\documentclass[preprint,epj]{svjour}
\usepackage{hyperref}
\usepackage{graphics}
\usepackage[super,compress]{cite}
\usepackage{amsmath}
\usepackage{graphicx}
\usepackage{float}
\usepackage{grffile}
\newcommand{\numberthis}{\addtocounter{equation}{1}\tag{\theequation}}
\begin{document}
\title{Finch-Skea Solutions of Anisotropic Stellar Models in $f(R)$ Gravity}

\author{B. Thakore \thanks{lightyear1998@gmail.com}\inst{,1}
\and R. Goti \thanks{rid181198@gmail.com}\inst{,1}
\and S. Shah \thanks{shlok.shah2223@gmail.com}\inst{,1}
\and H. Pandya \thanks{pandyahardey@gmail.com}\inst{,2} 
\and D. M. Pandya \thanks{dishantpandya777@gmail.com}\inst{,3}
}

\institute{Department of Physics, Pandit Deendayal Petroleum University, Gandhinagar-382007, Gujarat, India
\and {Department of Information and Communication Technology, Pandit Deendayal Petroleum University, Gandhinagar-382007, Gujarat, India}
\and Department of Mathematics, Pandit Deendayal Petroleum University, Raisan, Gandhinagar 382 007, India 
}

%
%
%
\date{Received: date / Revised version: date}
%
\abstract{
Present paper deals with the composition and modelling of compact dense astrophysical bodies under the framework of $f(R)$ gravity. The model is employed on various observed strange stars viz., SMC X-1, SAX J1808.4-3658, Swift J1818.0-1607, PSR J1614-2230 and PSR J0348+0432. Upon setting the appropriate value of dimensionless coupling parameter $\lambda $, the physical parameters such as the density, the radial and tangential pressures were obtained. Mass-Radius relation without presuming any equation of state is capable enough to accommodate all strange stars nearly having solar mass up to 2.5. The physical viability of the model is examined for all the aforementioned stars and it is found that all the regularity and stability conditions are satisfied. 
} 

\authorrunning {Thakore \emph{et al.}}
\titlerunning {F-S Solu. of Aniso. Ste. Models in F(R) Grav.}
\maketitle
\section{Introduction}	
The phenomenon of accelerated expansion of the universe was first noticed in 1998 while observing Type Ia Supernovae, in tandem with the discoveries of the Cosmic Microwave Background (CMB). The work done by eminent researchers such as Perlmutter \textit{et al.}\cite{Perlmutter1999}, Hawkins \textit{et al.}\cite{Hawkins2003} and Eisentein \textit{et al.}\cite{Eisentein2005} was instrumental in this paradigm-shifting discovery. This particular breakthrough paved the way for the existence of an exotic energy component called Dark Energy. Backed by numerous astronomical observations, dark energy has been a subject of intense research and debate in recent years. As per the conditions of the $\Lambda$CDM model, dark energy is essentially the Cosmological Constant, with the density accounting for 72$\%$ of the global energy budget of the universe. The remaining 28$\%$ is confined to normal matter in the form of galactic clusters (only 4$\%$), with the remaining 24$\%$ is occupied by cold dark matter (CDM), the nature of which remains unclear. To explain the occurrence of dark energy, theoretical physicists have often propounded the use of modified theories of gravity brought about by making changes to the Einstein- Hilbert action integral, leading to multiple theoretical possibilities, some of which include the $f(R)$, $f(\mathcal{T})$, $f(T)$ and $f(R, T)$ theories of gravity, where $R$ represents the scalar curvature, $\mathcal{T}$ represents the trace of the energy-momentum tensor, and $T$ represents torsion.\par
For more than four decades, modeling of neutron stars based on anisotropic parameters as opposed to isotropic composition has begun to gain prominence. Ruderman\cite{Ruderman1972} and Canuto\cite{Canuto1974} backed the idea of the existence of orthogonal pressure components, $p_r$ and $p_t$, where $p_r$ is the radial pressure component, while $p_t$ signifies the tangential component. R. F. Sawyer\cite{R.F.Sawyer1972} and A. I. Sokolov\cite{A.I.Sokolov1980} hypothesized that the said anisotropy may arise as a result of the presence of a Type III-A superfluid, a solid core at the center of the star, phase transitions, or pion condensations inside a neutron star. The modelling of such stars has been carried out in an extensive fashion through the use of general relativity. The exact mathematical solutions generated from these have a wide range of astrophysical applications. The modeling proved the anisotropy of stars by separating the radial and tangential pressure components. Mak and Harko\cite{MakHarko2004} provided a class of exact solutions of Einstein's field equations having an anisotropic source. Herrera \textit{et al.}\cite{Herrera2008} studied multiple static spherically symmetric solutions of Einstein's field equations and discussed the physical validity and implications of such solutions. Rahaman \textit{et al.}\cite{Rahaman2012} applied the Krori-Barua solutions\cite{KroriBarua75} to anisotropic compact charged stars. Sharma and Ratanpal\cite{SharmaRatanpal2013} provided a quadratic equation of state for neutron stars using the Finch-Skea ansatz\cite{FinchSkea1989}. Pandya \textit{et al.}\cite{Pandya2015} generated exact solutions of a generalized form of the Finch-Skea ansatz, and later on provided a way to generalize the Finch-Skea metric \cite{Pandya2020}. While compact star models have been described in substantial detail through the utilization of general relativity, further elaboration to these models and their related parameters can be propounded upon in terms of the inclusivity of dark energy based phenomena.
\par
This connotes the necessity of the said cosmological constant when considering the modelling of the universe as an ensemble of stars, black holes, planets and various other entities, as well as the modelling of individual dense binaries. We use the $f(R)$ gravity model for our purpose, more specifically, the Starobinksy model\cite{starobinsky1980} $f(R)=R+\Lambda R^2$.  Egeland\cite{Egeland2007} investigated that the cosmological constant would exist due to the density of the vacuum, this is a consequence of modeling the mass and radius of the neutron star. To demonstrate the validity of his assumption, Egeland used the relativistic equations of hydrostatic equilibrium with the fermion equation of state (EOS). The exact solutions for physically viable theoretical models for several dense stellar systems such as binary neutron star systems, X-ray bursters, and gamma-ray bursters have been described using the aforementioned model. \par

BC Paul \textit{et al.}\cite{ Paul2011}  modelled compact star described by Vaidya--Tikekar metric. Maharaj and Takisa\cite{Maharaj2012} also modelled compact star having anisotropic pressure and they also took electric-charge effects into consideration. For a specific polytropic index, exact solutions to Einstein’s field equations for an anisotropic sphere
admitting a polytropic EOS have been obtained by
Thirukkanesh and Ragel\cite{Thirukkanesh2012}. Hossein \textit{et al.}\cite{Hossein2012} developed anisotropic star models in the presence of a varying cosmological constant. M.H. Murad\cite{Murad2013}examined various charge distributions inside self-bound stars and provided analytical solutions to Einstein-Maxwell field equations in a metric ansatz suggested by Durgapal\cite{Durgapal1982}.  
Bhar \textit{et al.}\cite{Bhar2015} investigated the higher dimensional compact star. This work has been extended by Maurya \textit{et al.}\cite{Maurya2014} for the charged anisotropic compact stars. Ovalle\cite{Ovalle2017} provided the method of gravitational decoupling as a method of solving the field equations to model compact dense stars. K.N. Singh \textit{et al. } \cite{KNSingh2017} given model for compact stars whose interior admits Karmarkar Condition. Abbas \textit{et al.}\cite{Abbas2015} provided a compact star model for strange quintessence stars using the concept of $f(R)$ gravity. Variations to the Starobinsky model of $f(R)$ gravity has also been proposed by Astashenok \textit{et al.}\cite{Astashenok2017} as a way of incorporating the more exotic forms of matter. Sharif \textit{et al.}\cite{Sharif2019} described an alternative method for modeling the same using gravitational decoupling. Astashenok \textit{et al.}\cite{ Astashenok2014} provided a way to resolve the hyperon conundrum in $f(R)$ gravity, and provided another model \cite{Astashenok2019}, which included a way to model supermassive neutron stars in axion $f(R)$ gravity. In all of the mathematical models proposed using modified theories of gravity, we have seen extensive use of the Krori-Barua metric. \par
In this paper, we investigate the usage of the Finch-Skea metric in $f(R)$ gravity. In Section \ref{EH Action}, we elaborate upon the modification of the Einstein-Hilbert action in $f(R)$ gravity and in Section \ref{Staroexpression} Starobinsky's  $f(R)$ gravity model has been incorporated. Section \ref{PhysicalParameters} deals with the various physical parameters of the model. Sections \ref{CAnalysis} and \ref{GAnalysis} include subsequent conditional analysis and graphical analysis of the model. Inspection of the obtained \textit{M-a} plots is given in Section \ref{MaAnalysis} while the scope of the model and its possible applications have been highlighted in Section \ref{Secdiscuss}.

\section{Anisotropic Matter Configuration in \textit{f(R)} Gravity \label{EH Action}}
The modification of Einstein-Hilbert action in \textit{f(R)s} gravity can be done as:
\begin{equation}
\label{eqn1}
I = \int  dx^4\sqrt{-g}[f(R) + \mathcal{L}_{matter}],
\end{equation} 
where $g$ is the determinant of the metric tensor, $f(R)$ is any arbitrary function of $R$, the Ricci scalar and $\mathcal{L}$ is the Lagrangian density.
The variation of (\ref{eqn1}) with respect to the metric $g_{\mu\nu}$ leads to the following field equations:
\begin{equation}
\label{eqn2}
G_{\mu\nu} = R_{\mu\nu}-\frac{1}{2}Rg_{\mu\nu} = T_{\mu\nu}^{(matter)} + T_{\mu\nu}^{(curv)}
\end{equation}
where $T_{\mu\nu}^{(matter)}$ is the matter stress energy tensor scaled by ${\frac{1}{f'(R)}}$ and $T_{\mu\nu}^{(curv)}$ is the benefaction arising from the curvature due to the effective stress-energy tensor given as:
\begin{equation}
\label{eqn3}
T_{\mu\nu}^{(curv)} = \frac{1}{F(R)}\left[\frac{1}{2}g_{\mu\nu}(f(R) - RF(R)) + F(R)^{;\mu\nu}g_{\mu\alpha}g_{\nu\beta} - g_{\mu\nu}g_{\alpha\beta}\right],
\end{equation}
where, $F(R) = f'(R)$, i.e., derivative with respect to Ricci scalar $R$.
We select the spherically symmetric metric line element as:
\begin{equation}
\label{eqn4}
ds^2 = e^{\mu(r)}dt^2 - e^{\nu(r)}dr^2 - r^2(d\theta^2 + \sin^2{\theta}d\phi^2)
\end{equation}	

Considering 4-dimensional space-time to be a hyperspace, can be embedded into 5-dimensional bulk. This was conceptualized by the Randall-Sundrum (RS) second brane-world model \cite{Randall99}. It was found that an $n$-dimensional space can be embedded in $(n+k)$-dimensional space. Here $k$ is minimum number of extra dimensions and $V_n$ is called to be embedding 
class $k$ of $n$-dimensional space. Class-I type is demonstrated by Schwarzschild's interior solutions and the Friedmann Universe. For a spacetime to be of class-I, it must satisfy Karmarkar's condition \cite{Karmarkar48}. It is a necessary condition; however, it is not a self-sufficient parameter when discussing the physical viability of the compact celestial body in question. This condition can be obtained for charged as well as uncharged stellar models. It can be given as follows:
\begin{equation}
\label{eqn5}
R_{1414} = \frac{R_{1212}R_{3434} + R_{1224}R_{1334}}{R_{2323}}
\end{equation} 
Here, $R_{2323} \neq 0$ and $R_{abcd}$ are the non zero components of Riemann tensor. 

Using condition (\ref{eqn5}), the line element (\ref{eqn4}) can be altered to give the following differential equation:
\begin{equation}
\label{eqn6}
\frac{\mu'\nu'}{1 - e^\nu} = -2(\mu'' + {\mu'}^2) + {\mu'}^2 + \nu'\mu'
\end{equation} 

Solving equation (\ref{eqn6}), we get,
\begin{equation}
\label{eqn7}
e^\mu = (A + B\int \sqrt{e^\nu - 1}dr)^2
\end{equation}

The standard anisotropic fluid of the compact stars can be given as:
\begin{equation}
\label{eqn8}
T_{\alpha\beta}^m = (\rho + p_t)u_{\alpha}u_{\beta} - p_tg_{\alpha\beta} + (p_r - p_t)v_{\alpha}v_{\beta},
\end{equation}
where $u_{\alpha} = e^{\frac{\mu}{2}}{\delta_{\alpha}^0}$, $v_{\alpha} = e^{\frac{\nu}{2}}{\delta_{\alpha}^0}$ and $\rho$, $p_r$ and $p_t$ correspond to matter energy density, radial and transverse pressures respectively.
Thus the field equations can be modified with respect to spacetime as given in equation (\ref{eqn4}) as:
\begin{equation}
\label{eqn9}
\rho = \frac{e^{-\nu}}{2r^2}\left(r^2F'{\nu}' + 2Fr{\nu}' - fr^2e^{\nu} + Fr^2e^{\nu}R + 2Fe^{\nu} - 2r^2F'' - 4rF' - 2F\right)
\end{equation}

\begin{equation}
\label{eqn10}
p_r = -\frac{e^{-\nu}}{2r^2}\left(-2F + 2Fe^{\nu} - fr^2e^{\nu} + Fr^2e^{\nu}R - 2Fr{\mu}' - 4rF' - r^2F'{\mu}'\right)
\end{equation} 

\begin{equation}
\label{eqn11}
p_t = \frac{e^{-\nu}}{4r}\left(2Fa{\mu}'' - Fr{\nu}'{\mu}' + 2r{\mu}'F' + Fr{\mu}'^2 + 2F\mu - 2r{\nu}'F' + \psi(r) \right)
\end{equation} 
where $\psi(r)=- 2F{\nu}'
+ 2rfe^{\nu} - 2rFe^{\nu}R + 4rF'' + 4F'$.

Accordingly, the  Ricci scalar $R$ is given as:
\begin{equation}
\label{eqn12}
R = \frac{e^{-\nu(r)} \left(2 r^2 \mu''(r)+r^2 \mu'(r)^2+r \mu'(r) \left(4-r \nu'(r)\right)-4 r \nu'(r)-4 e^{\nu(r)}+4\right)}{2 r^2}
\end{equation}
\section{Constructing Anisotropic Star Models Using $f(R)$ Gravity  \label{Staroexpression}}

Now, to create a model of a compact star, the metric potential $e^\nu$ is assumed to be $1 + Cr^2$, as suggested by Finch and Skea \cite{FinchSkea89}. 

From equation (\ref{eqn7}), we get $\mu(r) = \log\left[\left(A + \frac{1}{2}Br\sqrt{r^2C}\right)^2\right]$, where, $A$ and $B$ are constants. Values of $A$, $B$ and $C$ are derived using the relevant boundary conditions.

Upon substituting $\nu(r)=\log(1 + Cr^2)$ and $\mu(r)=\log\left[\left(A + \frac{1}{2}Br\sqrt{r^2C}\right)^2\right]$, where $A$, $B$ and $C$ are constants, in equation (\ref{eqn4}), the expression for the metric changes to:
\begin{equation}
\label{FSMetric}
ds^2 = (A + \frac{1}{2}Br\sqrt{r^2C})^{2}dt^2 - (1+Cr^{2})dr^2 - r^2(d\theta^2 + \sin^2{\theta}d\phi^2)
\end{equation}\\
Accordingly, the Ricci Scalar as defined in \ref{eqn12} can be written as:\\
\begin{equation}
\label{eqn13}
R = \frac{2 \left(2 A C \left(C r^2+3\right)+B \sqrt{C} \left(C r^2-3\right) \left(C r^2+2\right)\right)}{\left(C r^2+1\right)^2 \left(2 A+B \sqrt{C} r^2\right)}
\end{equation}

For the early universe, there are a variety of inflationary models developed using scalar fields, which stem from super-gravity theories. Starobinsky \cite{starobinsky1980} suggested one of the earliest inflation models which corresponds to the conformal deviation in quantum gravity, essentially accounting for the presence of the expanding universe. The $f(R)$ gravity model given by Starobinsky is therefore represented as:
\begin{equation}
\label{eqn14}
f(R)= R + \lambda R^2
\end{equation}
where, $\lambda$ is a constant. Starobinsky proposed this constant as a way to elaborate upon the exponential growth found in the early time cosmological expansion. 

For this model, therefore, equations (\ref{eqn9}),(\ref{eqn10}) and (\ref{eqn11})become:
\begin{align*}
\rho = &\frac{1}{(C r^2+1)^5 (2 A+B \sqrt{C} r^2)^3} \cdot C (8 A^3 (2 C \lambda ( (C r^2 (-C^2 r^4+C r^2+ 45)- 69) \\ & +  (C r^2+3) (C r^2+1)^3) + 4 A^2 B \sqrt{C} (3 r^2 (C r^2+1)^3 (C r^2+3)-2 \lambda  (3 C r^2 \\&(C r^2-7)  +32)  (C r^2 (C r^2+6)-3))   + 2 A B^2 (2 \lambda  (C r^2 (C r^2 (C r^2 (-3 C^2 r^4\\&+3 C r^2+55)-303) +316)+108)+   3 C r^4 (C r^2+1)^3 (C r^2+3))\\&+  B^3 \sqrt{C} r^2 (-2 C \lambda  r^2 (C r^2 (C r^2 (C r^2 (C r^2\\ &- 1)+67)+341)+36)+C r^4 (C r^2+1)^3 (C r^2+3)+24 \lambda )) ] \numberthis  \label{eqn15}
 \end{align*}
\begin{align*}
 p_r = & \frac{1}{(C r^2+1)^4 (2 A+B \sqrt{C} r^2)^3}	\cdot 8 A^3 C (2 C \lambda  (C r^(C r^2+10)+37)-(C r^2+1)^3) \\ & + 4 A^2 B \sqrt{C} (2 C \lambda  (C r^2 (3 C r^2 (C r^2+10)+103)-64)-(C r^2+1)^3 (3 C r^2-4))\\ & +2 A B^2 C (r^2 (2 C \lambda  (C r^2 (3 C r^2 (C r^2+10)+47)-240)-(C r^2+1)^3 (3 C r^2-8))\\ &  -72 \lambda )  +B^3 C^{3/2} r^2 (2 \lambda  (C r^2-6) (C r^2 (C r^2 (C r^2+16)+45)+14)\\ &  -r^2 (C r^2-4) (C r^2+1)^3)
	\numberthis  \label{eqn16}
 \end{align*}
\begin{align*}
	p_t =  & \frac{1}{ (C r^2+1 )^5  (2 A+B \sqrt{C} r^2 )^3} [8 A^3 C (-2 C \lambda   (C r^2  (C r^2  (C r^2+17 )+91 )-37 )\\& - (C r^2+1 )^3 )+4 A^2 B \sqrt{C}  (-2 C \lambda   (3 C r^2  (C r^2  (C r^2  (C r^2+13 )+55 )-85 )\\&+64 )- (C r^2-4 )  (C r^2+1 )^3 )+2 A B^2 C  (r^2  (C r^2+1 )^3  (C r^2+8 )\\&-2 \lambda   (C r^2  (C r^2  (C r^2  (3 C r^2  (C r^2+9 )+41 )-431 )+148 )+36 ) )+B^3 C^{3/2} r^2  \\&(r^2  (2 C \lambda   (44-C r^2  (C r^2+5 )  (C^2 r^4-65 ) )+ (C r^2+4 )  (C r^2+1 )^3 )+24 \lambda  )]
	\numberthis  \label{eqn17}
\end{align*}
The anisotropy measurement $\Delta = \frac{2}{r} (p_t - p_r)$ for this model is given by: 
\begin{align*}
\Delta = &\frac{1}{ (C r^2+1 )^5  (2 A+B \sqrt{C} r^2 )^3} \cdot 2 C^{3/2} r  (8 A^3 \sqrt{C}  ( (C r^2+1 )^3-4 C \lambda   (C r^2  (C r^2\\&+14 )+69 ) )+4 A^2 B  ( (C r^2+1 )^3  (3 C r^2-2 )-4 C \lambda   (C r^2  (3 C r^2  (C r^2+12 )\\&+149 )-108 ) )+2 A B^2 \times \sqrt{C}  (r^2  ( (C r^2+1 )^3  (3 C r^2-4 )-4 C \lambda   (C r^2  (3 C r^2 \\& (C r^2+10 )+59 )-312 ) )+256 \lambda  )+B^3  (4 \lambda   (C r^2  (C r^2  (316-C r^2  (C r^2 \\& (C r^2+8 )-53 ) )+192 )+48 )+C r^4  (C r^2-2 )  (C r^2+1 )^3 ) )
\numberthis  \label{eqnanisotropy}
\end{align*}
$\Delta$ represents the anisotropy of compact star. The local anisotropy of fluid inside the star represents equivalent measurement of force. $\Delta>0$ is a repulsive behaviour while $\Delta<0$ is an attractive behaviour of fluid. When $p_t > p_r$, the anisotropy will be positive ($\Delta>0$). It is found that anisotropy is positive means the massive compact structure is stabilized with repulsive force and gravitational collapsing. At center of the star, both components of pressure will become equal which means $\Delta=0$.

\section{Physical Parameters of Compact Stars\label{PhysicalParameters}}
The connotations obtained from the field equations provide various parameters which are helpful in validating the model that has been obtained. This section discusses the physical parameters which are essential to the claim of a model being physically viable. 
\subsection{Density and Pressure Gradients}
The derivative of Equations (\ref{eqn15})-(\ref{eqn17}) with respect to the radial parameter will give,
\begin{align*}
\frac{d \rho}{dr} = \quad & \frac{1}{ (C r^2+1 )^6  (2 A+B \sqrt{C} r^2 )^4}-\frac{2 C^2 r  (C r^2+5 )}{ (C r^2+1 )^3} [ 8 C^{3/2} \lambda  r  (16 A^4 C^{3/2}  (C r^2 \\& (C r^2  (C r^2-3 )-89 )+195 ) +32 A^3 B C  (C r^2  (C r^2  (C r^2  (C r^2-3 )-77 )\\&+211 )-66 )+8 A^2 B^2 \sqrt{C}  (C r^2  (C r^2  (C r^2  (3 C r^2  (C r^2-3 )-179 )+689 )\\&-532 )-128 )+8 A B^3  (C r^2  (C r^2  (C r^2  (C r^2  (C r^2  (C r^2-3 )-25 )+279 )\\&-353 )-202 )-39 )+B^4 \sqrt{C} r^2  (C r^2  (C r^2  (C r^2  (C r^2  (C r^2  (C r^2-3 )+135) \\&+819 )+108 )-24 )-12 ) ) ]
\numberthis  \label{rhodr}
\end{align*}
\begin{align*}
\frac{d p_r}{dr} = \quad &\frac{1} { (C r^2+1 )^5  (2 A+B \sqrt{C} r^2 )^4} \cdot [2 C r  ( (C r^2+1 )^3  (2 A+B \sqrt{C} r^2 )^2  (4 A^2 C\\&+4 A B \sqrt{C}  (C r^2-2 )+B^2  (C  \times r^2  (C r^2-8 )-4 ) )+\lambda   (-64 A^4 C^2  (C r^2 \\& (C r^2+14 )+69 )-128 A^3 B C^{3/2}  (C r^2  (C r^2  (C r^2+14 )+66 )-31 )\\&-32 A^2 B^2 C  (C r^2  (C r^2  (3 C r^2  (C r^2+14 )+163 )-264 )-24 )\\&-32 A B^3 \sqrt{C}  (C r^2  (C r^2  (C r^2  (C r^2  \times (C r^2+14 )+18 )-219 )-59 )-3 )\\&+4 B^4 C r^2  (C r^2  (C r^2  (640-C r^2  (C r^2  (C r^2+14 )-107 ) )+380 ) \\& +84 ) ) )]
\numberthis  \label{prdr}
\end{align*}
\begin{align*}
\frac{d p_t}{dr}=\quad &\frac{1} { (C r^2+1 )^6  (2 A+B \sqrt{C} r^2 )^4} \cdot 4 C r ((C r^2+1)^3 (2 A+B \sqrt{C} r^2)^2 (4 A^2 C\\&+2 A B \sqrt{C} (C r^2-3)-B^2(C r^2 (C r^2+6)+2))+\lambda (32 A^4 C^2 (C r^2\\& (C r^2(C r^2+24)+165)-138)+64 A^3 B C^{3/2} (C r^2(C r^2(C r^2\\& (C r^2+21)+129)-183)+58)+16 A^2 B^2 C (C r^2 (C r^2 (C r^2(3 C r^2 (C r^2+17)\\&+241)-759)+364)+56)+16 A B^3 \sqrt{C}(C r^2 (C r^2(C r^2 (C r^2 (C r^2+1) (C r^2+11)\\&-372)+181)+78)+15)+2 B^4 C r^2 (C r^2 (C r^2 (C r^2 (C r^2 (C r^2-9) (C r^2+15)-780)\\&-132)-64)-12)))
\numberthis \label{ptdr}
\end{align*}

$\frac{d\rho}{dr} \leq 0$, $\frac{dp_r}{dr} \leq 0$, and $\frac{dp_t}{dr} \leq 0$, within region of our candidate star. These put bound on the model.

\subsection{Stability Expressions}
The stability analysis needs certain parameters to analyze the model. The parameters are written exhaustively in this Section. \\ \\
\textbf{(1) Radial and Tangential Sound Speed} \\
This includes radial and tangential sound speed for the analysis. The radial and tangential sound speed are denoted by $\upsilon_r$ and $\upsilon_t$, respectively.   The square of radial and transverse speeds are given as:
\begin{align*}
{\upsilon_r}^ 2 = {\frac{dp_r}{d\rho}} = &\{2 C r  ( (C r^2+1 )^3  (2 A+B \sqrt{C} r^2 )^2  (4 A^2 C+4 A B \sqrt{C}  (C r^2-2 )+B^2  (C r^2 \\& (C r^2-8 )-4 ) )+\lambda   (-64 A^4 C^2  (C r^2  (C r^2+14 )+69 )-128 A^3 B C^{3/2}  (C r^2 \\& (C r^2  (C r^2+14 )+66 )-31 )-32 A^2 B^2 C  (C r^2  (C r^2  (3 C r^2  (C r^2+14 )+163 )\\&-264 )-24 )-32 A B^3 \sqrt{C}  (C r^2  (C r^2  (C r^2  \times (C r^2  (C r^2+14 )+18 )-219 )\\&-59 )-3 )+4 B^4 C r^2  (C r^2  (C r^2  (640-C r^2  (C r^2  (C r^2+14 )-107 ) )+380 )\\&+84 ) ) )\} / \{ (C r^2+1 )^5  (2 A+B \sqrt{C} r^2 )^4  (\{8 C^{3/2} \lambda  r  (16 A^4 C^{3/2}  (C r^2  (C r^2 \\& (C r^2-3 )-89 )+195 )+32 A^3 B C  (C r^2  (C r^2  (C r^2  (C r^2-3 )-77 )+211 )\\&-66 )+8 A^2 B^2 \sqrt{C} \times (C r^2  (C r^2  (C r^2  (3 C r^2  (C r^2-3 )-179 )+689 )-532 )\\&-128 )+8 A B^3  (C r^2  (C r^2  (C r^2  (C r^2  \times (C r^2  (C r^2-3 )-25 )+279 )-353 )\\&-202 )-39 )+B^4 \sqrt{C} r^2  (C r^2  (C r^2  (C r^2  (C r^2  (C r^2  (C r^2-3 )+135 )+819 )\\&+108 )-24 )-12 ) ) \} / \{ (C r^2+1 )^6  (2 A+B \sqrt{C} r^2 )^4\}-\frac{2 C^2 r  (C r^2+5 )}{ (C r^2+1 )^3} ) \}
\numberthis  \label{soundr}
\end{align*}
\begin{align*}
{\upsilon_t}^ 2 = {\frac{dp_t}{d\rho}} = &\{4 C r \ (\ (C r^2+1\ )^3 \ (2 A+B \sqrt{C} r^2\ )^2 \ (4 A^2 C+2 A B \sqrt{C} \ (C r^2-3\ )-B^2 \ (C r^2 \\&\ (C r^2+6\ ) +2\ )\ )+\lambda  \ (32 A^4 C^2 \ (C r^2 \ (C r^2 \ (C r^2+24\ )+165\ )-138\ )\\&+64 A^3 B C^{3/2} \ (C r^2 \ (C r^2 \ (C r^2 \times (C r^2+21\ )+129\ )-183\ )+58\ )\\&+16 A^2 B^2 C \ (C r^2 \ (C r^2 \ (C r^2 \ (3 C r^2 \ (C r^2+17\ )+241)-759\ )+364\ )\\&+56\ )+16 A B^3 \sqrt{C} \ (C r^2 \ (C r^2 \ (C r^2 \ (C r^2 \ (C r^2+1\ ) \ (C r^2+11\ )-372\ )\\&+181\ )+78\ )+15\ )+2 B^4 C r^2 \ (C r^2 \ (C r^2 \ (C r^2 \ (C r^2 \ (C r^2-9\ ) \ (C r^2\\&+15\ )-780\ )-132\ )-64\ )-12\ )\ )\ )\} / \{\ (C r^2+1\ )^6 \ (2 A+B \sqrt{C} r^2\ )^4 \\& \ (\{8 C^{3/2} \lambda  r \ (16 A^4 C^{3/2} \ (C r^2 \ (C r^2 \ (C r^2-3\ )-89\ )+195\ )+32 A^3 B C \\& \ (C r^2 \ (C r^2 \ (C r^2 \ (C r^2-3\ )-77\ )+211\ )-66\ )+8 A^2 B^2\times \sqrt{C} \ (C r^2 \ (C r^2 \\& \ (C r^2 \ (3 C r^2 \ (C r^2-3\ )-179\ )+689\ )-532\ )-128\ )+8 A B^3 \ (C r^2 \\&\ (C r^2 \  \times (C r^2 \ (C r^2 \ (C r^2 \ (C r^2-3\ )-25\ )+279\ )-353\ )-202\ )-39\ )\\&+B^4 \sqrt{C} r^2 \ (C r^2 \ (C r^2 \ \times (C r^2 \ (C r^2 \ (C r^2 \ (C r^2-3\ )+135\ )+819\ )+108\ )\\&-24\ )-12\ )\ )\} / \{\ (C r^2+1\ )^6 \ (2 A +B \sqrt{C} r^2\ )^4 \}-\frac{2 C^2 r \ (C r^2+5\ )}{\ (C r^2+1\ )^3}\ ) \}
\numberthis  \label{soundt}
\end{align*}\\
From Equations (\ref{soundr}) and (\ref{soundt}), we have the expression for Herrera's condition as:
\begin{align*}
{\upsilon_t}^ 2 - {\upsilon_r}^ 2 = &\{ (C r^2+1 )^3  (2 A+B \sqrt{C} r^2 )^2  (4 A^2 \sqrt{C}  (C r^2-1 )+4 A B  (C r^2-1 )^2\\&+B^2 C^{3/2} r^4  (C r^2-5 ) )+4 \lambda   (-16 A^4 C^{3/2}  (C r^2  (C r^2  (2 C r^2+39 )+248 )\\&-69 )-32 A^3 B C  (C r^2  (C r^2  (2 C r^2  (C r^2+18 )+209 )-148 )+27 )\\&-16 A^2 B^2 \sqrt{C}  (C r^2  (C r^2  (C r^2  (3 C r^2  (C r^2+16 )+223 )-430 )+38 )+16 )\\&-8 A B^3  (C r^2  (C r^2  (C r^2  (C r^2  (C r^2  (2 C r^2+27 )+43 )-573 )-97 )+16 )\\&+12 )+B^4 \sqrt{C} r^2  (C r^2  (C r^2  (C r^2  (C r^2  (228-C r^2  (2 C r^2+21 ) )+1527 )\\&+1152 )+528 )+96 ) )\} / \{4 \lambda   (-16 A^4 C^{3/2}  (C r^2  (C r^2  (C r^2-3 )-89 )+195 )\\&-32 A^3 B C  (C r^2 (C r^2  (C r^2  (C r^2-3 )-77 )+211 )-66 )-8 A^2 B^2 \sqrt{C}  (C r^2 \\& (C r^2  (C r^2  (3 C r^2  (C r^2-3 )-179 )+689 )-532 )-128 )-8 A B^3  (C r^2  (C r^2\\&  (C r^2  (C r^2  (C r^2  (C r^2-3 )-25 )+279 )-353 )-202 )-39 )-B^4 \sqrt{C} r^2 \\& (C r^2  (C r^2  (C r^2  (C r^2  (C r^2  (C r^2-3 )+135 )+819 )+108 )-24 )-12 ) )\\&+\sqrt{C}  (C r^2+1 )^3  (C r^2+5 )  (2 A+B \sqrt{C} r^2 )^4\} < 0
\numberthis  \label{herrera}
\end{align*}
The expression presented in Equation (\ref{herrera}) should follow the variation: $|\upsilon_t^2-\upsilon_r^2| \leq 1$\cite{Herrera2004}. Also, the individual square of sound speed of each part should be in range of 0 to 1. These expressions are necessary tools for the further analysis of the conditions related to them. \\

\textbf{(2) Relativistic Adiabatic Index} \\
The tangential and radial adiabatic index are expressed as $\Gamma_r = \frac{\rho + p_r}{p_r}\cdot \frac{dp_r}{d\rho}$, and $\Gamma_t = \frac{\rho + p_t}{p_t}\cdot \frac{dp_t}{d\rho}$. The adiabatic indices in terms of the radial and tangential components are required to verify the model. This is essential to the modelling process, because these expressions are related to the stability of the star. The higher the stability of the star, the higher are its adiabatic indices. 

Due to the complexity of the expressions of the two conditions, we have carried out the graphical analysis, hence used it to verify the model.

\subsection{Surface Redshift}
The surface redshift $Z_s$ is another essential bound for the model validation\cite{Bohmer2006}. The surface redshift is associated with the metric coefficient. The surface redshift is given by,

\begin{equation} 
\label{surfaceshift}
Z_s = -1 + \sqrt{1 + Cr^2}
\end{equation}

\subsection{Gravitational Redshift}
The gravitational redshift $Z_g$ is formulated as:

\begin{equation}
\label{gravitationalshift}
Z_g = -1 + \frac{1}{A+\frac{1}{2} B \sqrt{C} r^2}
\end{equation}

$Z_g$ should be a decreasing function of the radial parameter $r$ for our model to be well behaved. This is evident from Figure \ref{fig14}.
\section{Conditional Analysis \label{CAnalysis}}
Many conditions are analyzed in this section. Using various stringent bounds like strong energy condition, Herrera's condition, surface redshift, Buchdahl, and adiabatic index conditions, the model has passed through these bounds.
\subsection{Matching Conditions for the Constants}
Goswami et al. \cite{Goswami2014} provided the extra matching conditions that arose when considering stellar modelling in modified theories of gravity, showing that the constraints on the thermodynamic properties and stellar structure are purely mathematical. As per the text, the Schwarzschild solution is determined to be the better choice in the exterior region when considering matching conditions in stars. This is an indicator of the fact that general vacuum solutions do not exist in $f(R)$ gravity as they do in general relativity, therefore, the solutions exist primarily in terms of the star mass and the boundary surface $r=a$.
In the following section, we compare corresponding terms of our interior space-time Equation (\ref{eqn4}) to Schwarzschild's exterior space-time metric given as:
\begin{equation}
\label{SzExt}
ds^2 =  \left(1-\frac{2 M}{r}\right){dt}^2 -\frac{{dr}^2}{1-\frac{2 M}{r}} - r^2 \left(\text{$d\theta $}^2 + \text{$d\phi $}^2 \sin ^2\theta\right)
\end{equation}
on the surface at $r = a$. The $A$, $B$, and $C$ constants are found by standard conditions for the compact stars in relativity. The required number of conditions are three because of the three unknown constants. These three standard conditions are represented by coefficients of interior metric (\ref{eqn4}) and exterior metric (\ref{SzExt}) at boundary $r=a$ as shown below,
\begin{equation}
\label{matching1}
e^{\mu}_{(r=a)}  = \left(A + B\int \sqrt{e^\nu - 1}dr\right)^2_{(r=a)} = \left( 1 - \frac{2M}{r}\right)_{(r=a)} 
\end{equation}
\begin{equation}
\label{matching2}
e^{\nu}_{(r=a)}  = \left( e^{\log(1 + Cr^2)}\right) _{(r=a)}  = \left( 1 - \frac{2M}{r}\right) ^{-1}_{(r=a)}
\end{equation}
And last one is derivative of both time coefficients of interior and exterior space-time metric,
\begin{equation}
\label{matching3}
\left( \frac{\partial e^{\mu(r)}}{\partial r}\right) _{(r=a)} =  \left( \frac{\partial \left( 1 - \frac{2M}{r}\right) }{\partial r}\right) _{(r=a)}
\end{equation}

Using first two Equations (\ref{matching1}) and (\ref{matching2}), the final comparisons are given by,

\begin{equation}
\label{eqn31}
\frac{1}{a^2 C+1}=1-\frac{2 M}{a}
\end{equation}
\begin{equation}
\label{eqn32}
\left(\frac{1}{2} {Ba} \sqrt{a^2 C}+A\right)^2 = 1 - \frac{2 M}{a} 
\end{equation}
Using (\ref{eqn31}), we get,
\begin{equation}
\label{Cvalue}
C = \frac{-2M}{a^2(2M - a)}
\end{equation}
Thus, using Equations (\ref{matching1})-(\ref{matching3}) and (\ref{Cvalue}), values of $A$ and $B$ can be written as:
\begin{equation}
\label{Avalue}
A = \frac{-M}{\sqrt{a^2 - 2aM}} + \sqrt{1 - \frac{2M}{a}}
\end{equation}

\begin{equation}
\label{Bvalue}
B = \frac{\sqrt{M}}{\sqrt{2\cdot a^{{3}}}}
\end{equation}
Here, $a$ is the star radius and $M$ is mass of the dense star. All three unknown constants are expressed in terms of $M$ mass of the star, and $a$ radius of the star.

\subsection{Surface Redshift Condition}
The surface redshift $Z_s$ must be finite for $0 \leq Z_s \leq 5$ \cite{Bohmer2006}. As shown in Figure \ref{fig15}, our model satisfies this condition. The expression of the surface redshift is given in Equation \ref{surfaceshift}. In Table {\ref{table3} we have given the values of surface redshift for our chosen realistic stars at $r = a$. It is seen that the values obtained conform with the conditions provided with the values being less than $5$.
	
	\subsection{Buchdahl's Condition}\label{subsecbuchdahl}
As per Buchdahl\cite{Buchdahl1979}, the mass radius relationship for all realistic stellar objects must obey the inequality $\frac{M}{a} \leq \frac{4}{9}$. According to Table \ref{table4}, we can see that our model satisfies this inequality for all the five stars.

\subsection{Analysing the Bound on Model Constants}
We use the limits obtained for physical validity of the model to gauge the approximate bounds on the model constraints. The simplest of these bounds occur in the form of the surface redshift condition (\ref{surfaceshift}) and the Buchdahl conditions. The former condition helps us find the direct bounds on $C$, while the latter can be manipulated to derive the limits on $A$ and $B$.\\
From (\ref{surfaceshift}), we have:\\
\begin{equation}
\label{redshiftbounds}
0\leq -1+\sqrt{1+Cr^{2}} \leq 5
\end{equation}
This gives us 
\begin{equation}
\label{redshiftrearr}
0\leq Cr^{2} \leq 35
\end{equation}
At the surface, we take $r=a$, where $a$ denotes the radius of the star.
Therefore, using the surface redshift conditions, we get:
\begin{equation}
\label{CRedshift}
0\leq C \leq \frac{35}{a^2}
\end{equation}
Now, according to Buchdahl's condition,
\begin{equation}
\label{buchdahlboun1}
0<\frac{M}{a}\leq\frac{4}{9}
\end{equation}
Which gives us
\begin{equation}
\label{BuchdahlC1}
0<\frac{2M}{a^2}<\frac{8}{9a}
\end{equation}
In a similar fashion,
\begin{equation}
\label{BuchdahlC2}
0<\frac{1}{-2M+a}<\frac{9}{a}
\end{equation}
Multiplying Equation (\ref{BuchdahlC1}) with (\ref{BuchdahlC2}), 
\begin{equation}
\label{Buchdahlcombined}
0<\frac{2M}{a^{2}(-2M+a)}<\frac{8}{a^2}
\end{equation}
The expression obtained in the inequality is the same as the expression for $C$, as seen in (\ref{Cvalue}). Therefore, we have another bound for $C$, between $0<C<\frac{8}{a^2}$. Similar manipulations of the Buchdahl condition yield the limits of $A$ and $B$ as $\frac{-1}{9}<A<1 $ and $0<B<\frac{\sqrt{2}}{3a}$. \\
From the bounds obtained, we see the existence of radial dependencies on both $B$ and $C$, whereas $A$ seems to be independent of dependencies on the stellar radii. 
Therefore, considering the limits obtained from (\ref{CRedshift}) and the subsequent manipulations of the Buchdahl conditions, the final limits of the constants can be listed as:
\begin{eqnarray}
\label{limitsfinal}
\frac{-1}{9}<A<1\\
0<B<\frac{\sqrt{2}}{3a}\\
0<C<\frac{8}{a^2}
\end{eqnarray}

\begin{table}[ht]
	\centering
	\scriptsize
	\caption{Values of constants A, B and C for chosen stars:} 
	{
		\begin{tabular}{cccccc}\hline \\
			Star & M & $ a $   & A & B & C\\
			Name & ($ M_{\odot} $) & (km) &  &\\ \hline \\
			
			SMC X--1 & 1.04 & 8.301 &0.674944& 0.0367455 &  0.00860237 \\
			SAX J1808.4--3658(SS2) & 0.9 & 7.951 & 0.71154 & 0.0364646 & 0.00801291 \\
			Swift J1818.0--1607 & 2.0 & 12.5 & 0.56035 & 0.0275761 & 0.00579688 \\ PSR J1614--2230 & 1.908 & 13 & 0.605911  & 0.0253949 & 0.00457341 \\
			PSR J0348+0432 & 2.01 & 13 & 0.579189 & 0.0260655 & 0.00502597
		\end{tabular} \label{table1}}
\end{table}

Table \ref{table1} gives the values of $A$, $B$, and $C$ for five different stars. Mass and Radius of SMC X--1, SAX J1808.4--3658(SS2) are given by Rawls \textit{et al.} \cite{Rawls11} and  Elebert \textit{et al.} \cite{Elebert09}. The newly discovered magnetar Swift J1818.0--1607 is also considered, as given in \cite{NewMagnetar2020}. Mass and Radius of PSR J1614--2230 are proposed in Arzoumanian \textit{et al.} \cite{ Arzoumanian} and Demorest \textit{et al.} \cite{Demorest} respecively. Similarly, Mass and Radius of PSR J0348+0432 are taken from Antoniadis \textit{et al.} \cite{Antoniadis} and Zhao \cite{Zhao}. After putting mass and radius of the given stars, it is possible to calculate the numerical values of $A, B, C$ from Equations (\ref{Cvalue})-(\ref{Bvalue}), and derive the exact theoretical values of $\rho$, $p_r$ and $p_t$.

\subsection{Monotone Decrease Condition}
The condition suggests that derivative of the energy density, tangential pressure, and radial pressure should be in negative quadrant. Also, this condition represents that the density, and pressure will decrease with radial coordinate. This is called monotone decrease of these physical parameters ($\rho$, $p_r$, $p_t$). Therefore, the maximum values of density and pressure will be located at $r=0$. And the condition is given by,
\begin{center}
	\boldmath{$\frac{d\rho}{dr} \leq 0$, $\frac{dp_r}{dr} \leq 0$, and $\frac{dp_t}{dr} \leq 0$}
\end{center}
The expressions of $\frac{d\rho}{dr}$, $\frac{dp_r}{dr}$, and $\frac{dp_t}{dr}$ are given in Equations (\ref{rhodr})-(\ref{ptdr}). The profiles are in negative and decreasing with radial parameter. After certain point, the trends are increasing till the boundary of the stars. That means pressure, and density are decreased gradually towards the boundary after certain point.

\subsection{Central Anisotropy Condition}
The radial pressure and tangential pressure become equal at the center of the star. That means anisotropy which represents the difference between radial and tangential pressure becomes zero. This condition should be satisfied to avoid collapse at the center \cite{Pandya2020}. The condition is given by,
\begin{center}
	\boldmath{$\Delta_{(r=0)}=0$}
\end{center}
The $\Delta$ expression is given in Equation \ref{eqnanisotropy}. Also, the pictorial description is given in the Figure \ref{fig7} and $\Delta$ has null value at center for selected candidate stars. Also, the anisotropy is increasing radially. 

\subsection{Energy Conditions}
The energy bounds and strength of gravitational field are analyzed by several energy conditions and it is necessary to satisfy them. The validity of these energy conditions is vital for a physically reasonable energy-momentum tensor. The strength of gravitational field is described by energy conditions and this gravitational field should be enough to sustain the star's stability. The energy conditions for anisotropic fluid are defined by the following relations,
\begin{equation}
\label{eqnNEC}
\text{NEC}: \qquad \rho + p_r \ge 0, \quad \rho + p_t\ge 0 
\end{equation}
\begin{equation}
\label{eqnWEC}
\text{WEC}: \qquad \rho  \ge 0, \quad \rho - p_r \ge 0 \quad \rho - p_t \ge 0 
\end{equation}
\begin{equation}
\label{eqnSEC}
\text{SEC}: \qquad \rho - p_r - 2 p_t \ge 0 
\end{equation}
\begin{equation}
\label{eqnDEC}
\text{DEC}:\qquad   \rho > |p_r| \quad \rho > |p_t| 
\end{equation}

Here, NEC, WEC, SEC, and DEC stand for null energy condition, weak energy condition, strong energy condition, and dominant energy condition, respectively. Apparently, NEC will be satisfied because the profile of density, tangential, and radial pressure are in positive behaviour. WEC and SEC are described in Figures (\ref{fig8})-(\ref{fig10}). It is clear from the figures that conditions are appropriately satisfied. DEC is combination of NEC and WEC conditions. The energy condition analysis for the stars is shown in the Table \ref{table2}.

\begin{table}[ht]
	\centering
	\scriptsize
	\caption{Central and Surface Densities and Strong Energy Condition at  $r = a$ and $r = 0$ for chosen stars:}
	{\begin{tabular}{ccccccc}\hline \\
			Star & M & $ a $   & $\rho_c$ & $ \rho_s$ & $(\rho - p_r - 2p_t)_{r=0}$ & $(\rho - p_r - 2p_t)_{r=a}$ \\
			Name & ($ M_{\odot} $) & (km) & (MeV fm$^{-3}$) & (MeV fm$^{-3}$) & (MeV fm$^{-3}$)& (MeV fm$^{-3}$) \\ \hline \\
			SMC X--1 & 1.04 & 8.301 & 727.704 & 363.67 & 702.5087  & 337.804755  \\
			SAX J1808.4--3658(SS2) & 0.9 & 7.951 & 679.422 & 352.452 & 640.7078 & 313.7378\\
			Swift J1818.0--1607 & 2.0 & 12.5 & 504.598 & 187.192 & 324.1968 & 143.5138 \\
			PSR J1614--2230 & 1.908 & 13 & 400.283  & 164.814 & 282.812 & 130.692 \\
			PSR J0348+0432 & 2.01 & 13 & 439.164 & 169.865 & 295.502 & 132.069
		\end{tabular}\label{table2}} 
\end{table}

\begin{table}[htbp]
	\centering
	\scriptsize
	\caption{ Red-shift and Adiabatic Index at $r = a$ and $r = 0$ for chosen stars: }
	{
		\begin{tabular}{ccccc}\hline \\
			& M & $ a $  & $Z_s$ & $\Gamma_r$  \\
			Star Name & ($ M_{\odot} ) $ & (km) & Surface red-shift & Adiabatic-Index  \\ \hline \\
			
			SMC X--1 & 1.04 & 8.301 & 0.262045 & 2.55384  \\
			SAX J1808.4--3658(SS2) & 0.9 & 7.951 & 0.213592 & 2.66197  \\
			Swift J1818.0--1607 & 2.0 & 12.5 & 0.380493 & 1.94174 \\   PSR J1614--2230 & 1.908 & 13 & 0.331505  & 1.98721  \\
			PSR J0348+0432 & 2.01 & 13 & 0.359922 & 1.9423
			
		\end{tabular} \label{table3}}
\end{table}

\subsection{Stability Conditions Analysis}

\textbf{(1) Herrera's Cracking Condition} \\
Herrera\cite{Herrera2004} provided the cracking condition for a stable anisotropic compact star that results when equilibrium is disturbed which may be consequence of local anisotropy. This condition is constituted in terms of radial and tangential sound speed. $\upsilon_r$ and $\upsilon_t$ are radial and tangential sound velocity through the star. \\ 
Herrera's condition is given by,
\begin{center}
	\boldmath{$|\upsilon_t^2-\upsilon_r^2| \leq 1$}
\end{center}
Using Herrera's condition\cite{Herrera2004}, the stability of the star is extensively verified with Figure \ref{fig13}. Taking this condition into consideration, Abreu \textit{et al}\cite{abreu2007} demonstrated that parts of the sphere where $-1\leq \upsilon_t^2-\upsilon_r^2 \leq 0$ is potentially stable, while the parts where $0 < \upsilon_t^2-\upsilon_r^2 \leq 1$ is potentially unstable when $p_t$ is not zero within this sphere. It is noticeable from Table \ref{table4} that all the chosen star candidates for our model are compatible with the former condition and hence our model is potentially stable.

Also, the Equations (\ref{soundr})-(\ref{herrera}) represent the square of the radial and tangential sound speed. After deciding the parameters in Section \ref{PhysicalParameters}, the causality condition can be given by,
\begin{center}
	\boldmath{$0 \leq \sqrt{\frac{dp_r}{d\rho}} \leq 1$ and $0 \leq \sqrt{\frac{dp_t}{d\rho}} \leq 1$}
\end{center}

Here $\sqrt{\frac{dp_r}{d\rho}}$ and $ \sqrt{\frac{dp_t}{d\rho}}$ are $\upsilon_r$ (radial sound speed) and $\upsilon_t$ (tangential sound speed). The range of these two parameters are described in the Figures (\ref{fig11})-(\ref{fig12}). It is apparent that candidate stars lie between 0 and 1. This verified the stability of the stars using Causality condition. 

 These  conditions are needed to be satisfied. The probability of the gravitational collapse is much higher if the conditions are not compatible with the model. 

\textbf{(2) Relativistic Adiabatic Index} \\
The condition is,
\begin{center}
	\boldmath{ $\Gamma_r > \frac{4}{3}$}
\end{center}
The pictorial representaion of condition is shown in the Figure \ref{fig16} with reasonable conciliation. It shows that the profile increases exponentially after certain point but at center, the values are greater than 1.333 as per the condition. This is also evident from Table \ref{table3}.

	\begin{table}[ht]
		\centering
		\scriptsize
		\caption{Buchdahl Ratio and Herrera's Condition at $r=0$ and $r=a$ for chosen stars:}	
		{
			\begin{tabular}{ccccccc}\hline \\
				& M & $a$  & $\frac{M}{a}$ & $(\upsilon_t^2 - \upsilon_r^2)_{r=0}$ & $(\upsilon_t^2 - \upsilon_r^2)_{r=a}$ \\
				Star Name & ($ M_{\odot} ) $ & (km)  \\ \hline \\
				
				SMC X--1 & 1.04 & 8.301 & 0.18 & - 0.0115 & - 0.0819  \\
				SAX J1808.4--3658(SS2) & 0.9 & 7.951 & 0.17 & - 0.0182 & - 0.0845 \\
				Swift J1818.0-1607 & 2.0 & 12.5 & 0.24 & - 0.0330 & - 0.0578 \\  PSR J1614--2230 & 1.908 & 13 & 0.21 & - 0.0451 & - 0.0612 \\
				PSR J0348+0432 & 2.01 & 13 & 0.23 & - 0.0399
				& - 0.0586

			\end{tabular}\label{table4}}
	\end{table}

\section{Graphical Analysis \label{GAnalysis}}

In this section, we will try to inspect the similarity between the model provided here and results of empirical studies. We have used five strange star candidates SMC X--1, SAX J1808.4--3658 and Swift J1818.0--1607, PSR J1614--2230, PSR J0348+0432 with masses $1.04~M_{\odot}$, $0.9~M_{\odot}$ and $2.0~M_{\odot}$, $1.908~M_{\odot}$ and $2.01~M_{\odot}$ respectively, out of which Swift J1818.0--1607 is a magnetar. Taking value of speed of light, $c$ to be 299792 $\text{km}\cdot\text{s}^{-1}$ and universal gravitational constant $G$ to be $6.67430\cdot10^{-20} ~\text{km}^{-3}\cdot\text{kg}^{-1}\cdot\text{s}^{-2}$, values of constants $A$, $B$ and $C$ are calculated for all the stars and are given in Table \ref{table1}. The value of $\lambda$ is taken to be $0.35$. We check the variations of various physical parameters by choosing the given value of constants. \par
Figures \ref{fig1}, \ref{fig2} and \ref{fig3} demonstrate the dependence of matter density $\rho$, radial pressure $p_r$ and transverse pressure $p_t$ on radial distance $r$, which for all the five stars are monotonically decreasing in nature. This indicates compactness of each star at its centre. Figures \ref{fig4}, \ref{fig5} and \ref{fig6} represent density, radial pressure and transverse pressure gradient, which are less than zero, in the range $0 \leq r \leq a$, for the chosen stars. The anisotropy $\Delta$, variation for the five stars is plotted in Figure \ref{fig7}. It can be noticed that $\Delta > 0$ for both the stellar configurations, in the range $0 < r \leq a$. This generates a repulsive force, which acts as a counter phenomenon against gravitational collapse. Also, at $r = 0$, $\Delta = 0$, which makes our model well behaved. Figures \ref{fig8} and \ref{fig9} depict the weak energy conditions for stellar stability, and Figure \ref{fig10} depicts the strong energy condition. Both these conditions have positive values for the five chosen realistic stars in the range $0 < r \leq a$. Thus, $\rho$ is greater than $p_r$ and $p_t$. Figures \ref{fig11} and \ref{fig12} show that radial and transverse sound speed profiles lie in $[0,1]$ for all the selected stars. Figure \ref{fig13} illustrates that $|\upsilon^2_t - \upsilon^2_r|$ lies in $[0,1]$, which makes these stars stable in this theory. Figure \ref{fig14} shows gravitational redshift $Z_g$, which is a decreasing function of $r$, while Figure \ref{fig15} shows surface redshift $Z_s$, whose value lies in $[0,5]$, which provides increased credibilty to the claim of our model being stable. The adiabatic index in Figure \ref{fig16} exemplify that $\Gamma_r > \frac{4}{3}$ throughout the distribution of the chosen stellar bodies.\par
In the Krori-Barua metric, as shown by Abbas \textit{et al.}\cite{Abbas2015}, the EOS has a linear behaviour. A similar trend can be observed in the current model, wherein, it can be elicited from Figures \ref{fig17} and \ref{fig18} that EOS exhibits nearly linear behaviour for all the five stars. The graph is plotted between $\rho$ versus $p_r$, which demonstrates the linear nature of the EOS. Finally, the \textit{M-a} plots are described in Fig. \ref{fig19}  for both cases of $\rho_s=360~\text{MeV}\cdot \text{fm}^{-3}$ and for $\rho_s=450 ~\text{MeV}\cdot \text{fm}^{-3}$ at $\lambda= 0.35$. From the profiles for both cases, the trend looks like increasing with a negative rate. It is apparent from the curves that the trends in the case of $\rho_s=360 ~\text{MeV}\cdot \text{fm}^{-3}$ yield differing maximum and minimum mass values than the trend in the case of $\rho_s=450 ~\text{MeV}\cdot \text{fm}^{-3}$. Using these observations, the trend observed from the masses and their physical implications are given in the section below.  \par
The graphical analysis of stars of larger masses leads to intriguing trends being observed in the density and pressure gradients, wherein the curves are much shallower in Figures \ref{fig4}, \ref{fig5} and \ref{fig6} for the stars Swift J1808.0-1607, PSR J1614-2230 and PSR J0348+0432, when compared with the lower mass stars (SMC X-1 and SAX J1808.4-3685). Larger radii for the higher mass stars coupled with lower values of density and pressure parameters also indicates a decrease in the anisotropy. This decrease in pressure anisotropy also indicates softer equations of state. As shown in Figures \ref{fig17} and \ref{fig18}, this can be verified by the fact that the EOS obtained is more linear than low mass stars. \par
Another interesting property in the purview of the gravitational redshift as described in Figure \ref{fig14} emerges with respect to the hyperon connotations in maximal neutron star masses. PSR J0348+0432, when compared to the magnetar Swift J1808.0-1607, has similar mass and radius orientations. However, the gravitational redshift of the former is considerably lower than that of the latter. According to a treatise detailing the hyperon coupling constants and their relation with surface gravitational redshift in maximal mass neutron stars, the presence of the coupling constant for hyperons $\Xi$ reduces the gravitational redshift for neutron stars of similar masses \cite{XianFeng2019}, suggesting the presence of hyperon structures inside PSR J0348+0432, a hypothesis which has been proposed in the paper, and has been verified through the graph presented in the current paper. Finally, further analysis of the \textit{M-a} diagrams has been provided in the following section.  \\

\section{Analysis of the \textit{M-a} plots \label{MaAnalysis}}
It is of an imperative consequence to calculate the permissible masses for compact star models generated by various theories of gravity, lest the model admits masses of higher capacity and provides spurious values of the density and the pressure parameters, therefore leading to incorrect physical analyses of the stars in question. The \textit{M-a} plot provides the maximum and minimum possible mass at a given surface density $\rho_s$. Here, we have obtained \textit{M-a} for the Starobinsky model of $f(R)$ gravity. As described in the sections above, using the Starobinsky model  $\lambda = 0.35$, we compared the results of \textit{M-a} curve for different surface densities for the given metric. The \textit{M-a} profiles for both cases $\rho=360 ~\text{MeV} \cdot \text{fm}^{-3}$ and $\rho=450 ~\text{MeV} \cdot \text{fm}^{-3}$ have been provided in Fig.\ref{fig19}. Due to the complexity of the expressions, the usage of discretized values to obtain the \textit{M-a} plot was considered an appropriate measure. 
To compare the variations in the maximum and minimum masses which can be attained by stars at particular surface density, we described a tabular data for the outcome of our model. As shown below, the table elaborates upon the results of the \textit{M-a} plot:

\begin{table}[ht]
	\centering
	\scriptsize
	\caption{Maximum and minimum mass results for $\lambda=0.35$ at surface denisty of $\rho_s=360$ and $\rho_s=450 ~\text{MeV}\cdot \text{fm}^{-3}$}	
	{
			\begin{tabular}{cccc}\hline \\
			$\rho_s~(\text{Mev}\cdot \text{fm}^{-3})$	& $\text{Maximum Mass}$ ($M_\odot$) &$\text{Minimum Mass}$ ($M_\odot$ \\
			
			\hline \\
			
			$360$  & $M = 2.66$ &  $M=0.155$ \\
			$450$  & $M = 2.36$ &   $M=0.20$\\

		\end{tabular}\label{table5}}
\end{table}
In Fig. \ref{fig19} and Table \ref{table5}, it is observed that for $\rho_s= 360~\text{MeV} \cdot \text{fm}^{-3}$, the neutron star masses reach a maximum mass of 2.66 $M_\odot$ . When compared to the maximum mass of stars at $\rho_s= 450~\text{MeV} \cdot \text{fm}^{-3}$, we find that the mass of the star in question peaks at 2.36 $M\odot$.  When we observe the minimum masses, however, we see a different scenario. The minimum mass is higher for $450~\text{MeV} \cdot \text{fm}^{-3}$  surface density and lower for $360~\text{MeV} \cdot \text{fm}^{-3}$. The minimum masses at  $\rho_s= 450$ and $360\text{MeV} \cdot \text{fm}^{-3}$ are 0.20 and 0.155 $M_\odot$  respectively. However, it is also observed that for the lower masses and similar radii, the masses for $\rho_s= 450~\text{MeV} \cdot \text{fm}^{-3}$ exceeds, in value, the mass at $\rho_s= 360~\text{MeV} \cdot \text{fm}^{-3}$.  This indicates that the range of the aforementioned masses becomes progressively restricted at higher densities. 
\section{Discussion \label{Secdiscuss}}
Using the Finch-Skea ansatz, we provided a physically viable, analytical model of anisotropic compact celestial bodies. The solutions obtained supports the choice of Finch-Skea metric as a suitable candidate for modelling in $f(R)$ gravity. Using this model we can also infer the presence of hyperons in neutron stars with higher than usual maximal masses, specifically for the stars PSR J0438+0432 and PSR J1614-2230. The model verifies the presence of hyperon structures in this star through two primary factors: the presence of a soft EOS due to the linear nature of the pressure-density relation, and the relation between the presence of hyperon structures and their effect on gravitational redshift for compact stars of larger masses. Another factor which makes the Finch-Skea model a viable candidate for compact star modelling in $f(R)$ gravity is the fact that the model has been proven to be valid for a wide range of stars. This model is also valid for magnetars which is demonstrated here by evaluating it for newly discovered Swift J1818.0--1607 \cite{NewMagnetar2020}. The model also provides a generalized algebraic solution to the constants $A$, $B$ and $C$ in the Finch-Skea metric, and makes it suitable for use in further modelling endeavours. The pressures, as obtained above, can be described as combined forces, arising from myriad  phenomena such as neutron-hyperon interactions, phase transitions, electromagnetic interactions, etc. The paper and its concomitant  EOS can also be utilised to provide further knowledge about the quark-gluon interactions at the center of strange stars, which may provide information and variables which are necessary to glean more reasons behind a star’s inherent anisotropy. Another standout feature of the paper is that the \textit{M-a} plots were obtained solely through the usage of the analytical solutions, without the EOS as a pre-requisite towards obtaining the \textit{M-a} relations of the plot. This indicates that the \textit{M-a} plots hold true for any EOS of linear nature, which, in turn, has been proven through Figures \ref{fig17} and \ref{fig18}. A physically viable model for neutron stars has therefore been suitably obtained through the application of $f(R)$ gravity on the Finch-Skea metric. 

\begin{figure}[H]
	\centering
	\includegraphics[scale=0.5]{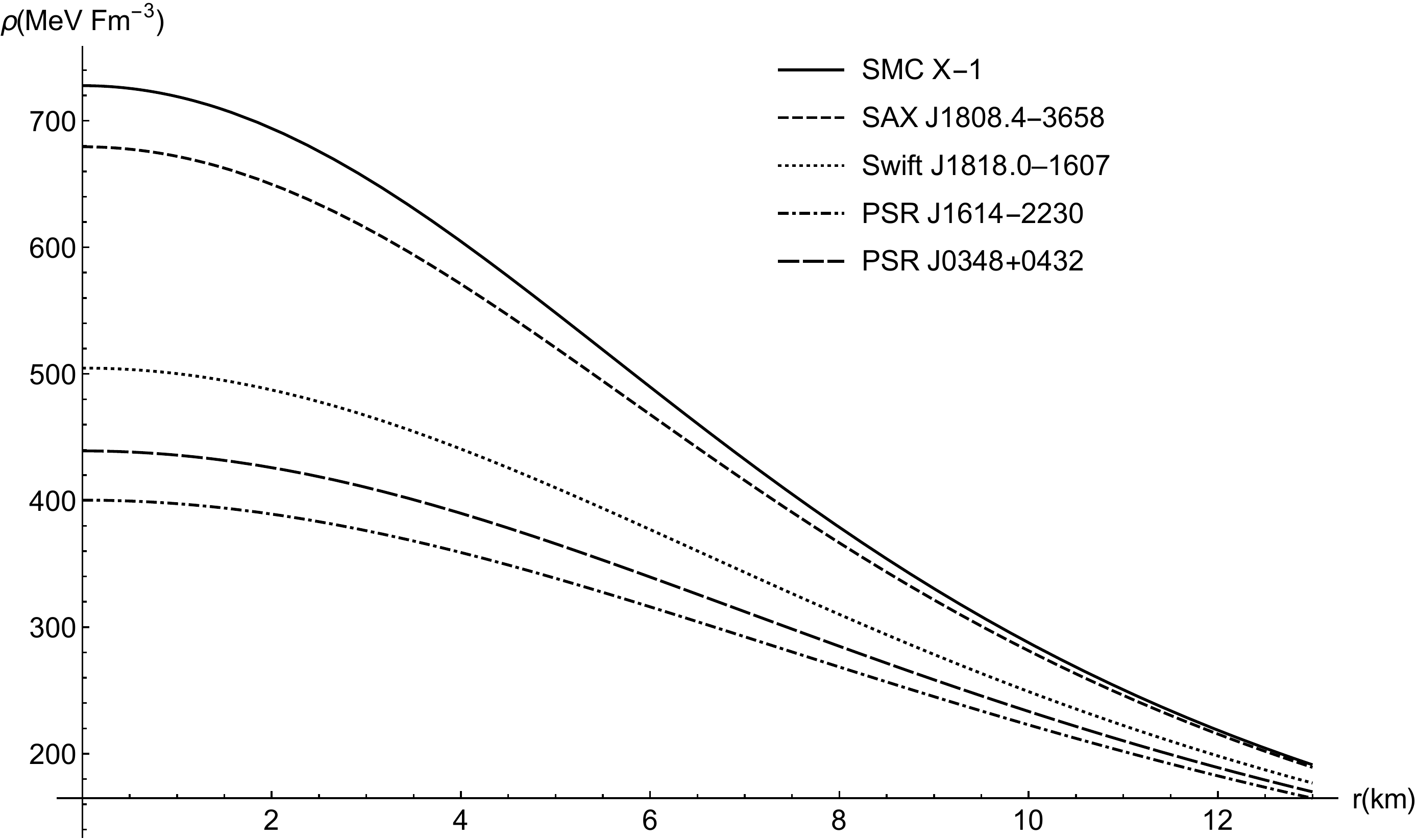}
	\caption{Density profile \label{fig1}}
\end{figure}

\begin{figure}[H]
	\centering
	\includegraphics[scale=0.5]{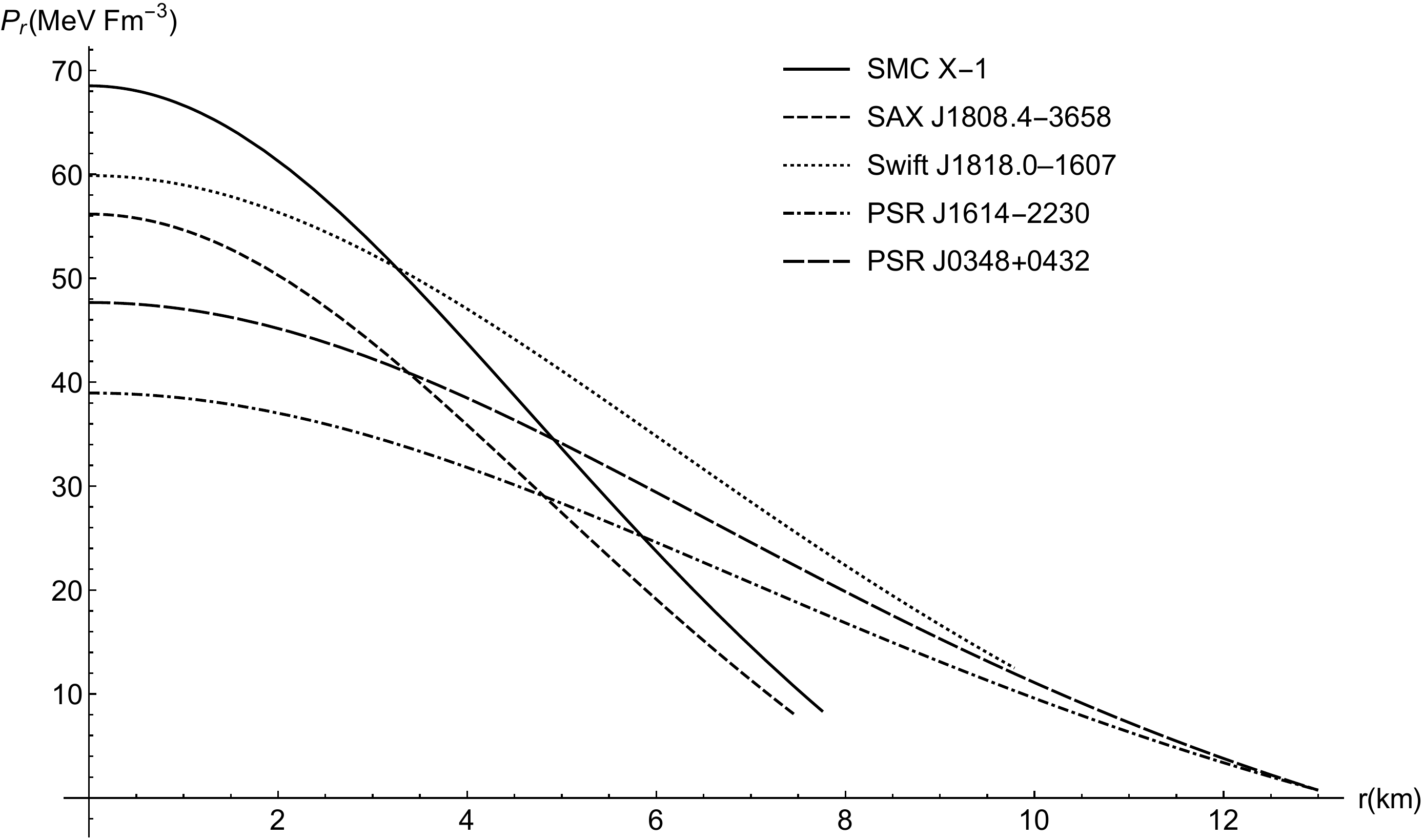}
	\caption{Radial Pressure profile \label{fig2}}
\end{figure}

\begin{figure}[H]
	\centering
	\includegraphics[scale=0.5]{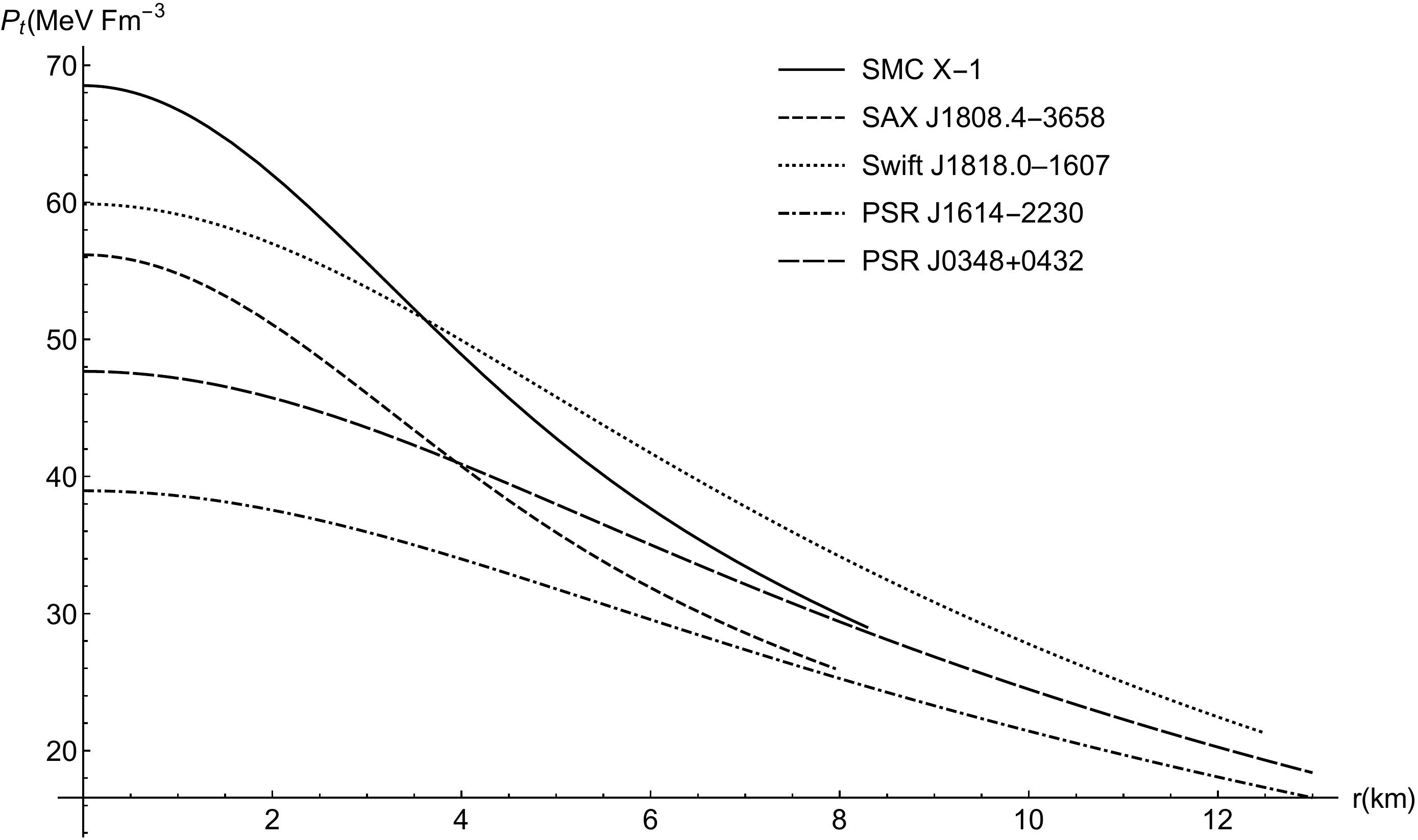}
	\caption{Transverse Pressure profile \label{fig3}}
\end{figure}

\begin{figure}[H]
	\centering
	\includegraphics[scale=0.5]{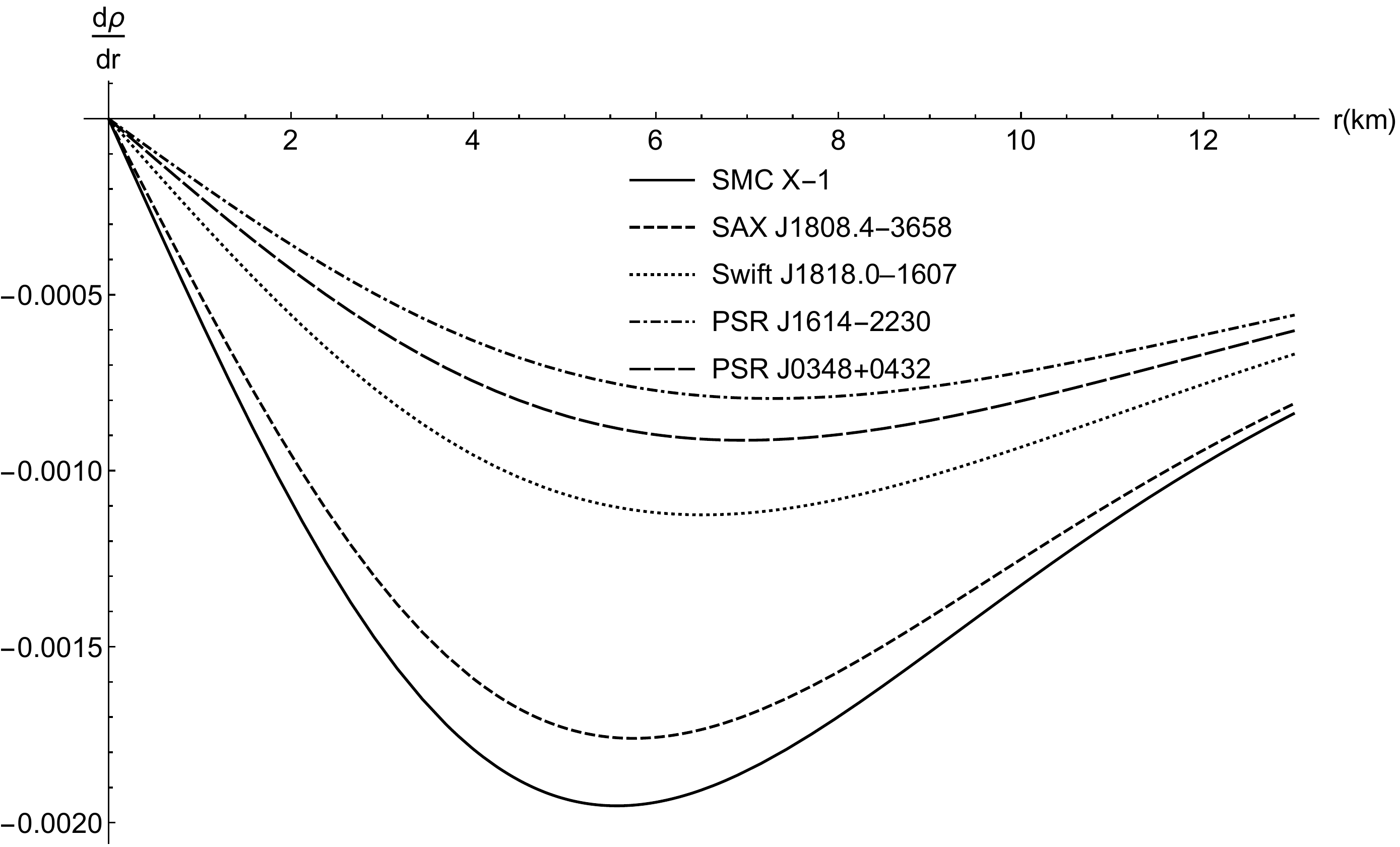}
	\caption{Density Gradient \label{fig4}}
\end{figure}

\begin{figure}[H]
	\centering
	\includegraphics[scale=0.5]{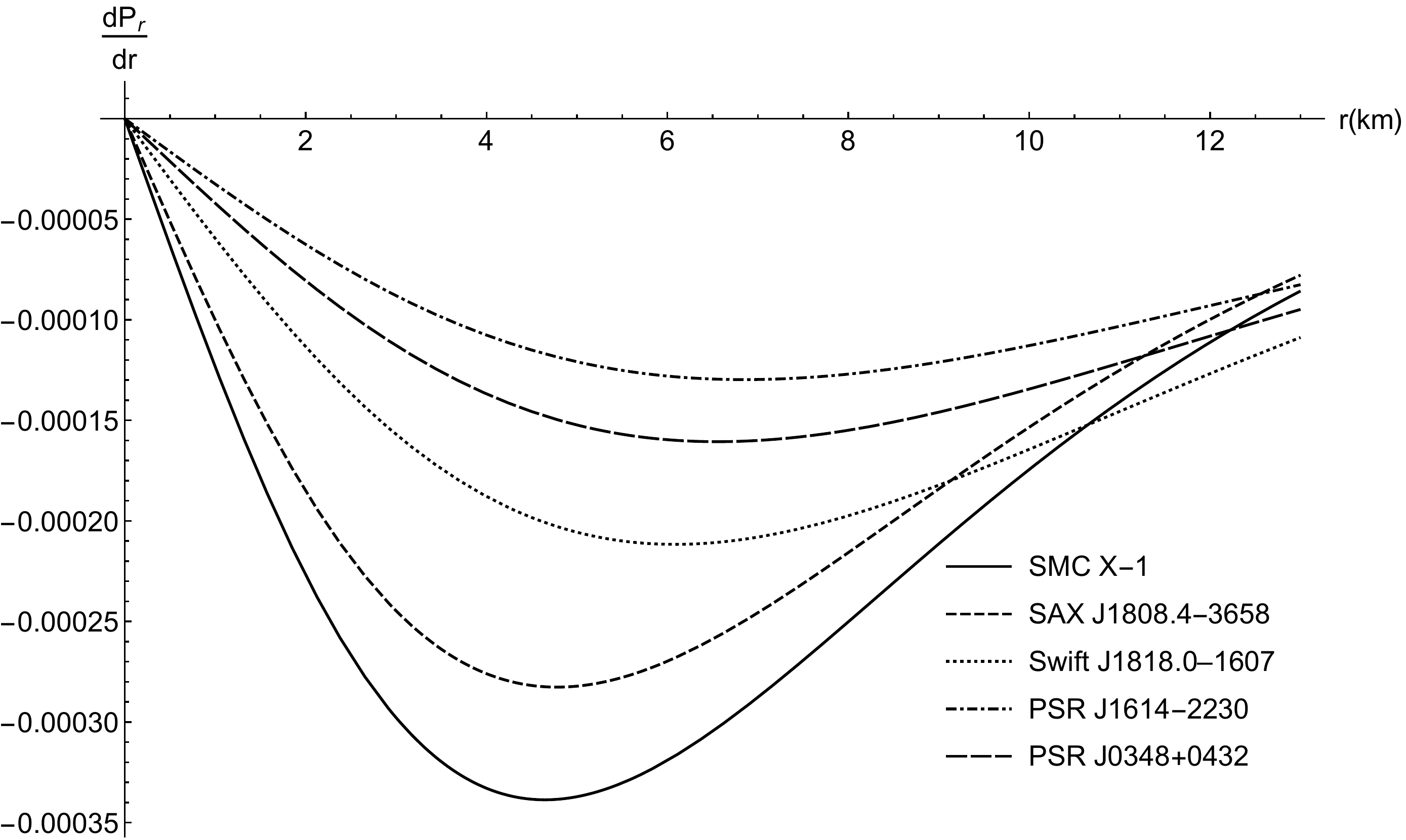}
	\caption{Radial Pressure Gradient \label{fig5}}
\end{figure}

\begin{figure}[H]
	\centering
	\includegraphics[scale=0.5]{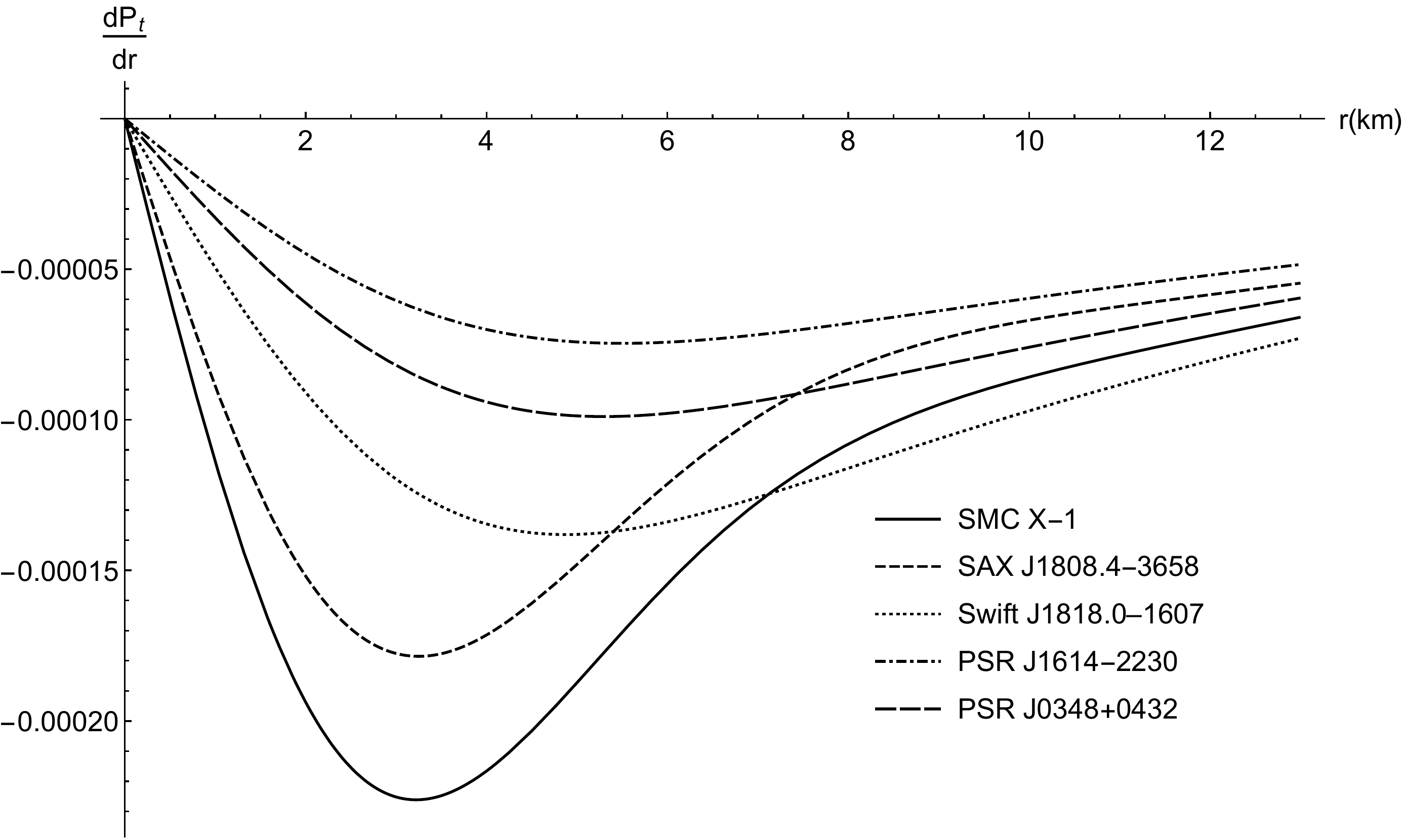}
	\caption{Transverse Pressure Gradient \label{fig6}}
\end{figure}

\begin{figure}[H]
	\centering
	\includegraphics[scale=0.5]{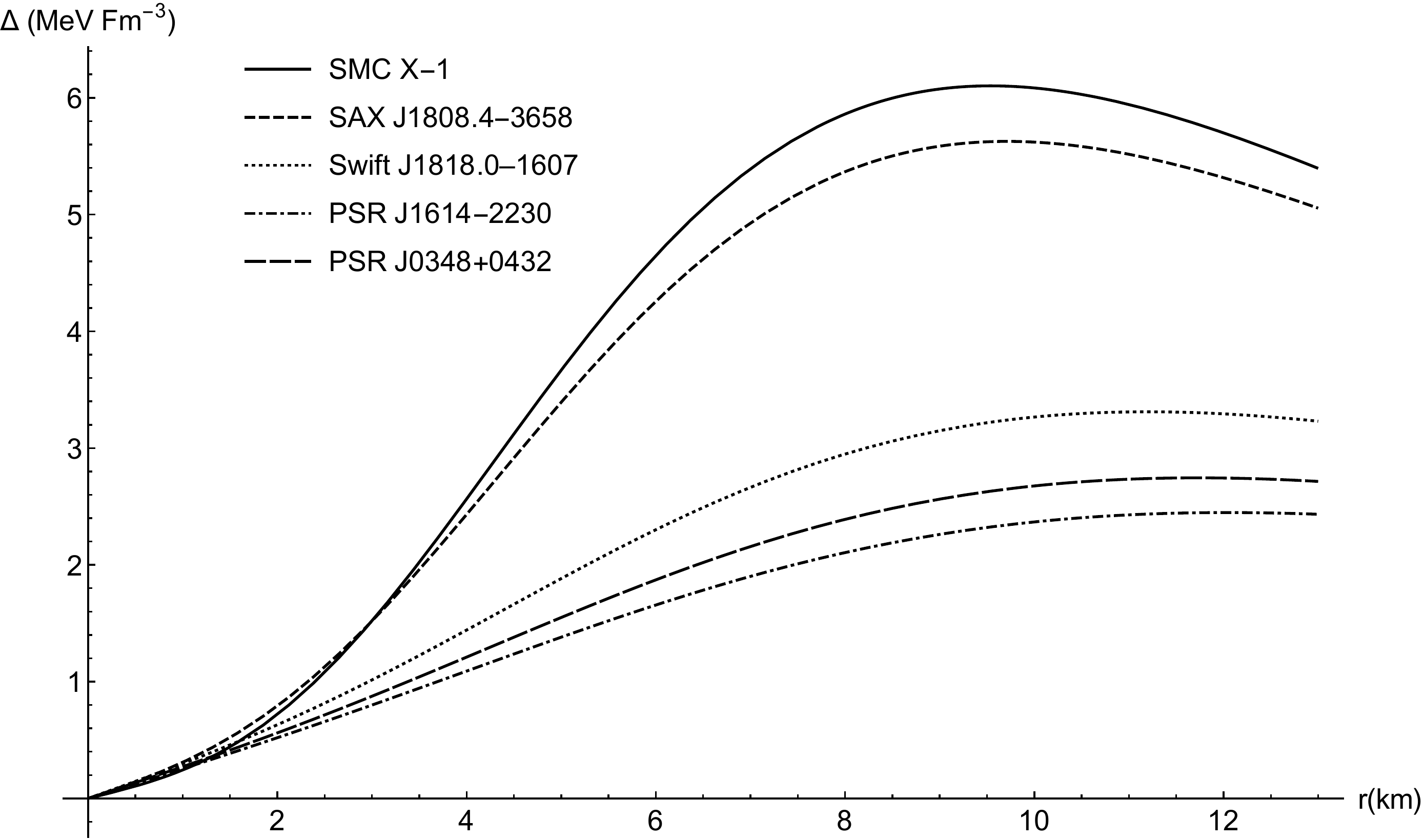}
	\caption{Anisotropy profile \label{fig7}}
\end{figure}

\begin{figure}[H]
	\centering
	\includegraphics[scale=0.5]{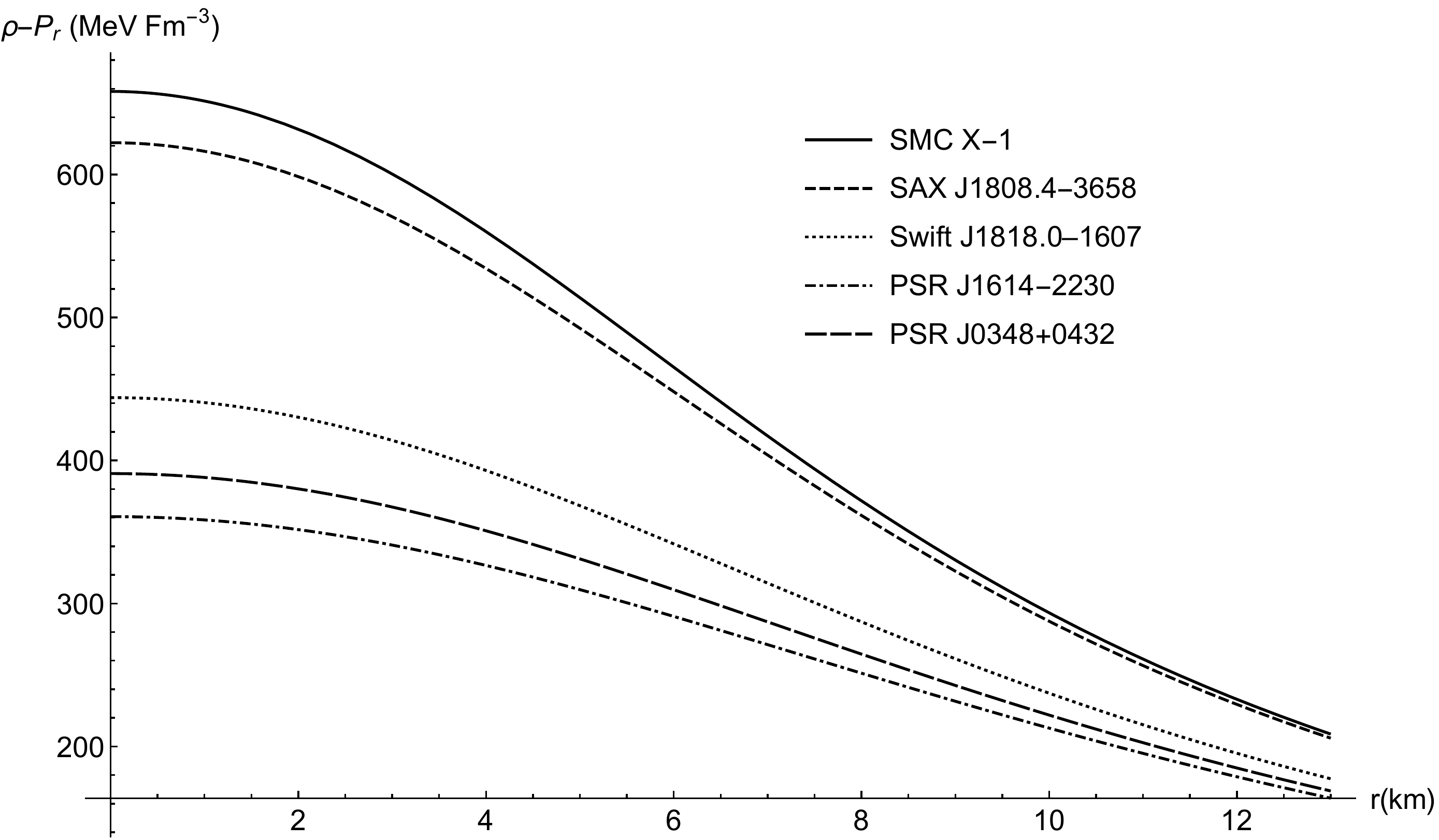}
	\caption{Weak Energy Condition profile $(\rho - p_r)$ \label{fig8}}
\end{figure}

\begin{figure}[H]
	\centering
	\includegraphics[scale=0.5]{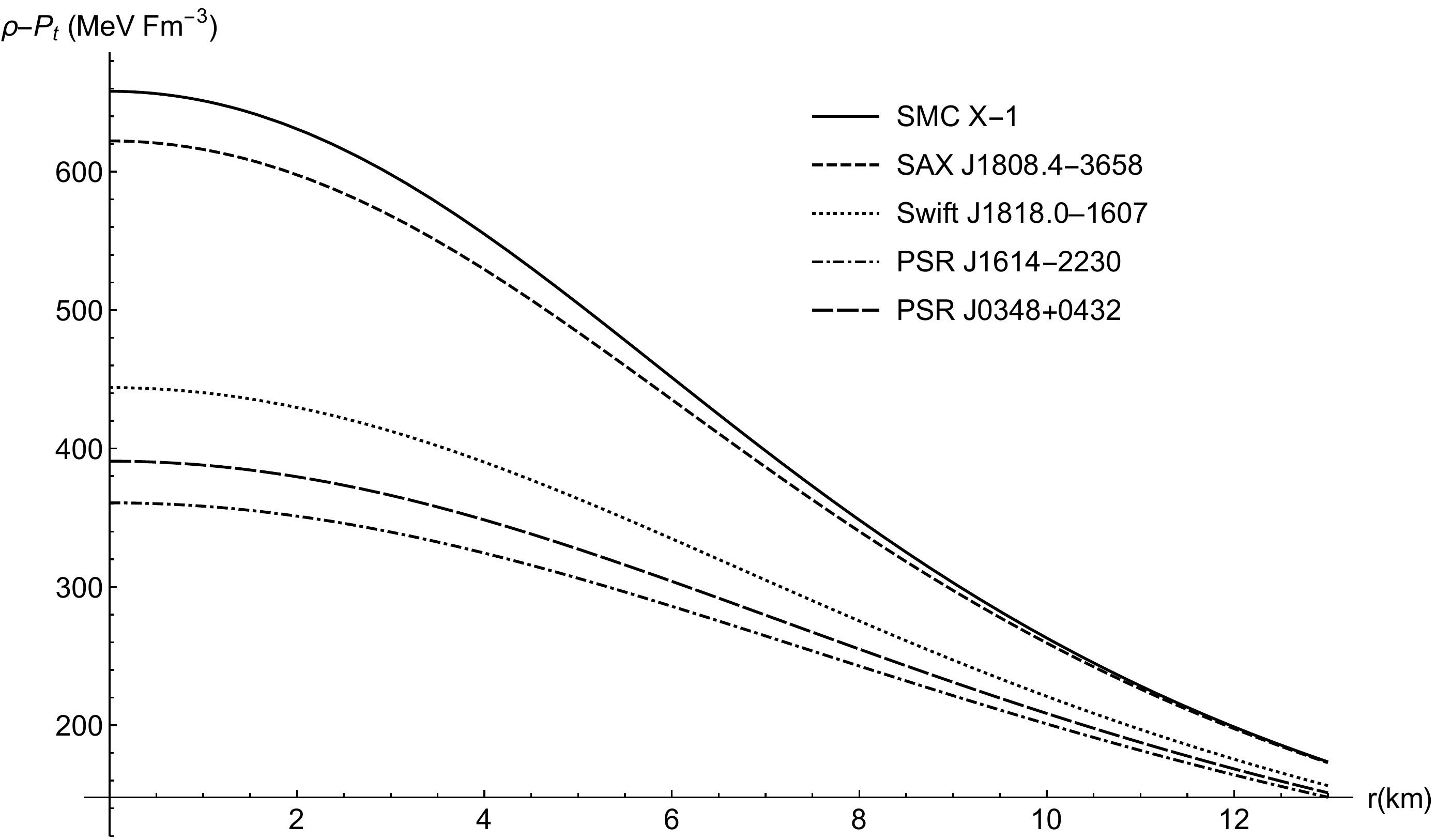}
	\caption{Weak Energy Condition profile $(\rho - p_t)$ \label{fig9}}
\end{figure}

\begin{figure}[H]
	\centering
	\includegraphics[scale=0.5]{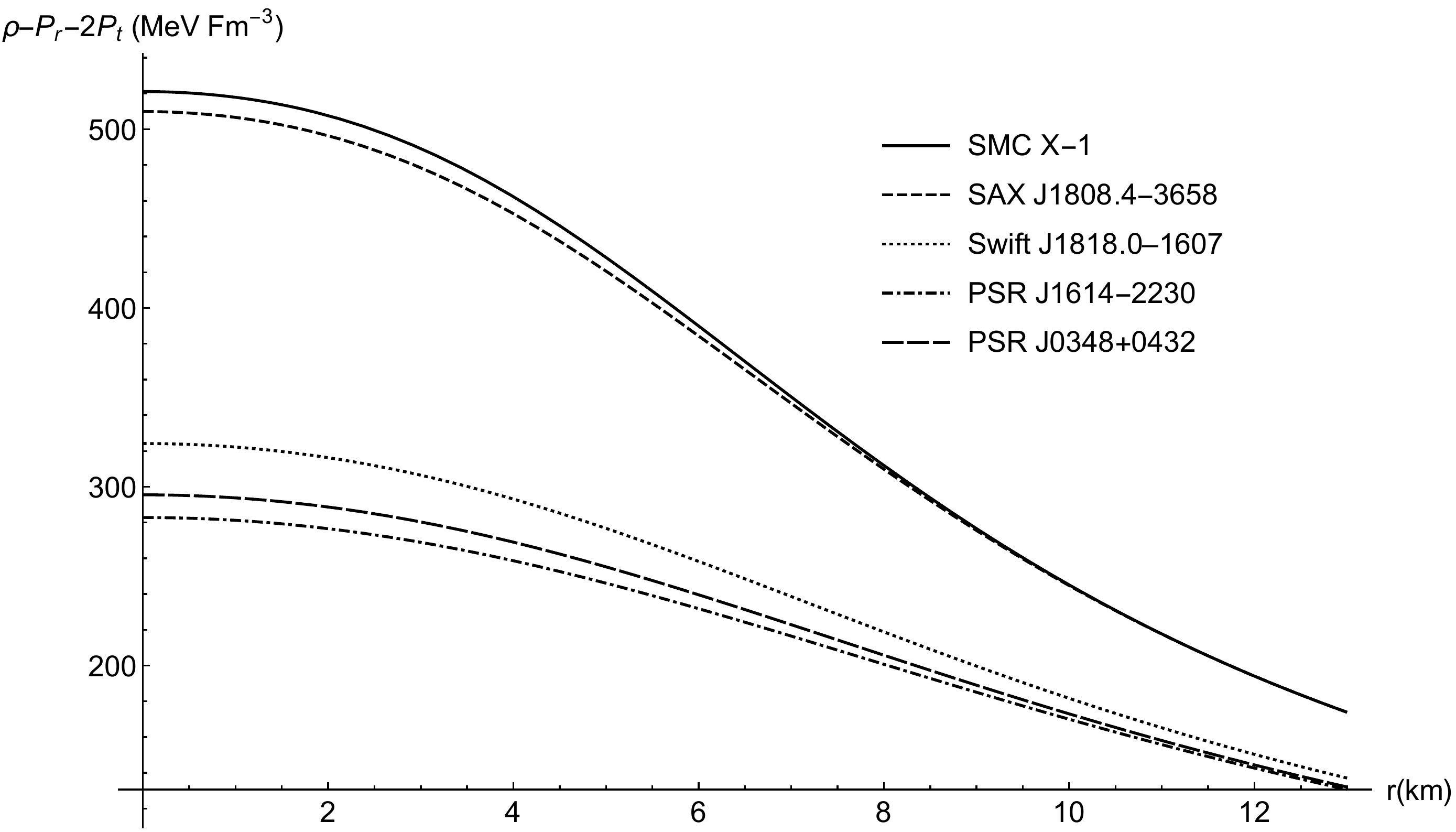}
	\caption{Strong Energy profile $(\rho - p_r - 2p_t)$ \label{fig10}}
\end{figure}

\begin{figure}[H]
	\centering
	\includegraphics[scale=0.5]{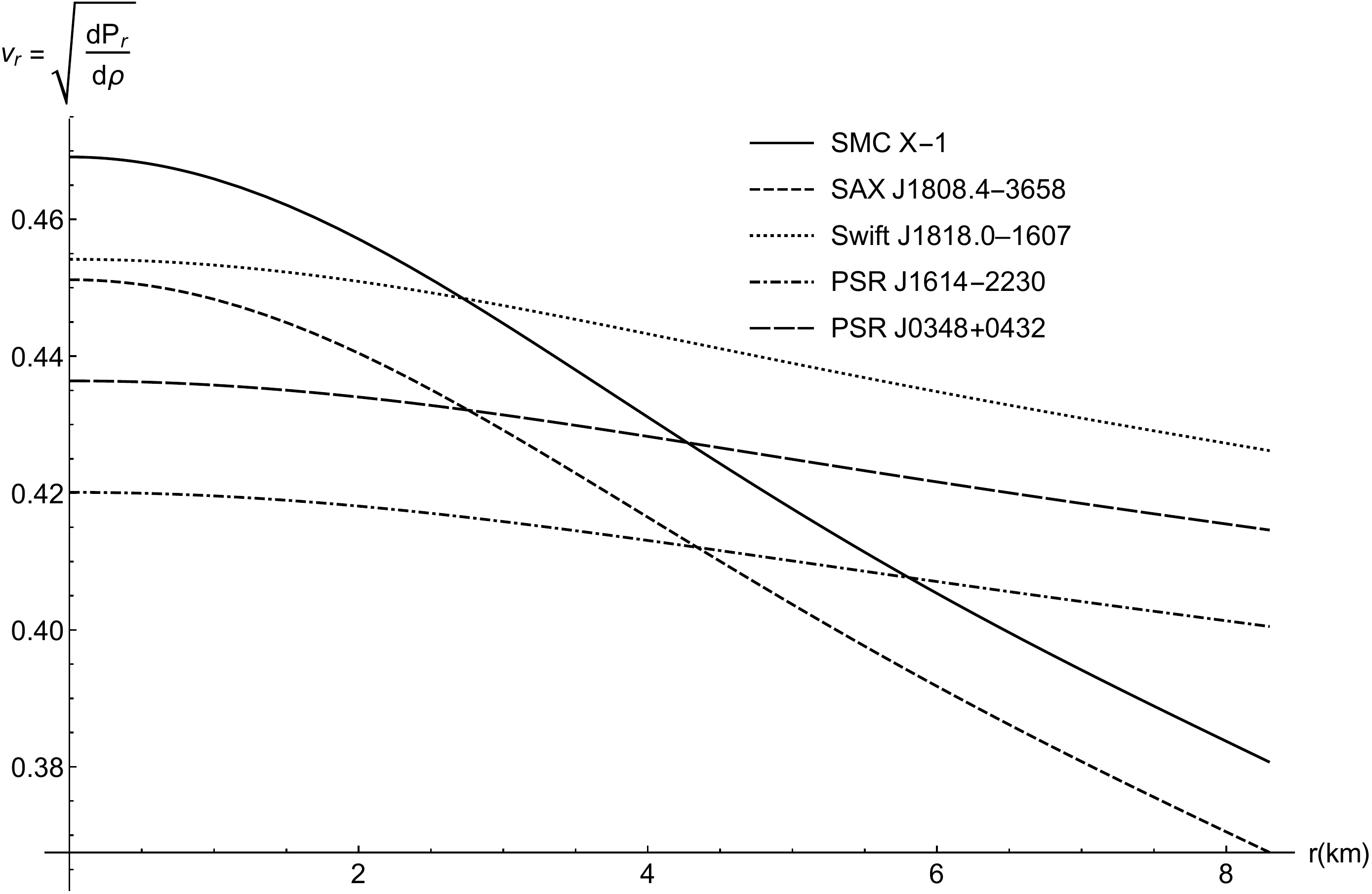}
	\caption{Radial Sound Speed profile \label{fig11}}
\end{figure}

\begin{figure}[H]
	\centering
	\includegraphics[scale=0.5]{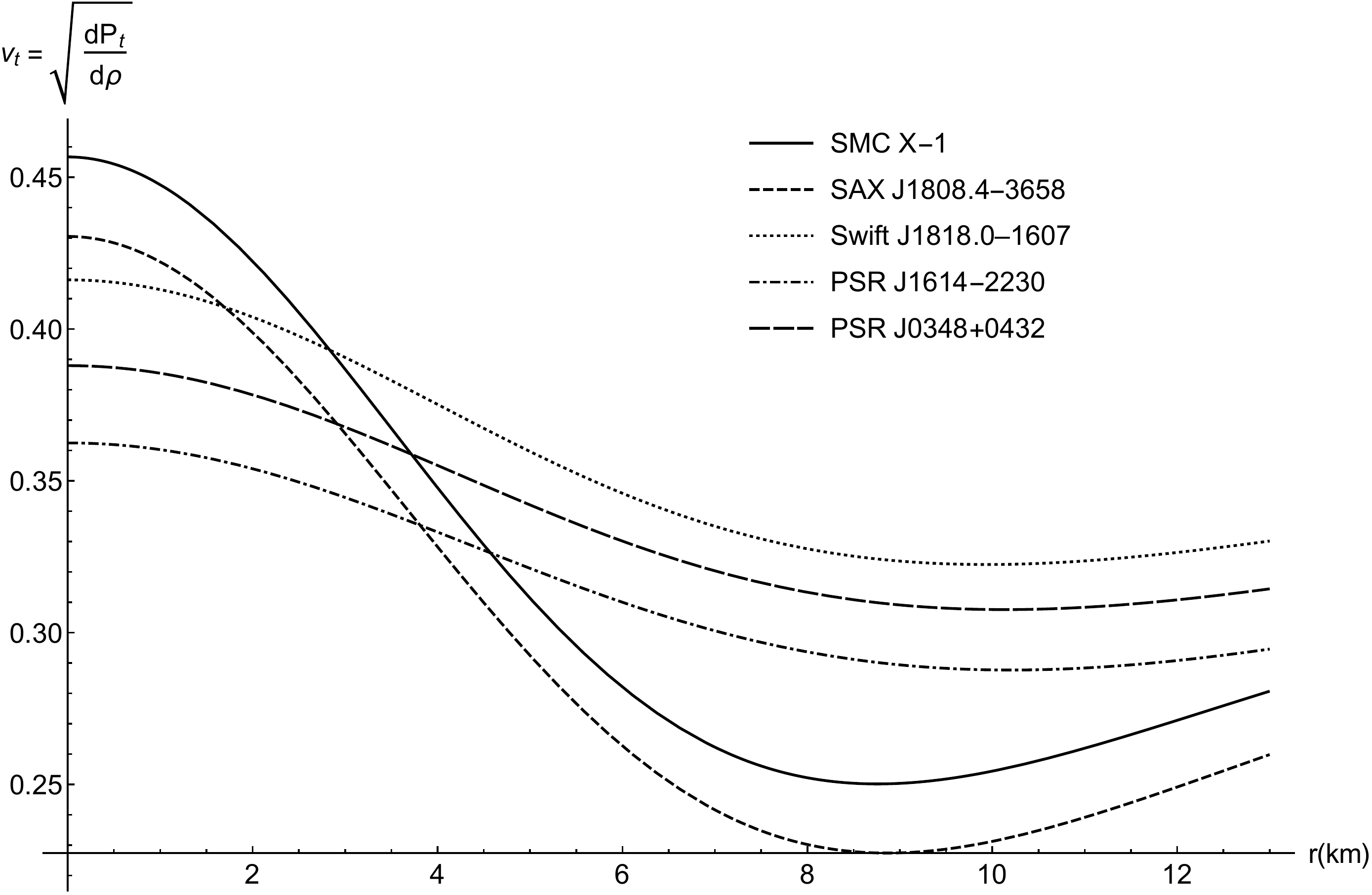}
	\caption{Transverse Sound Speed profile \label{fig12}}
\end{figure}

\begin{figure}[H]
	\centering
	\includegraphics[scale=0.5]{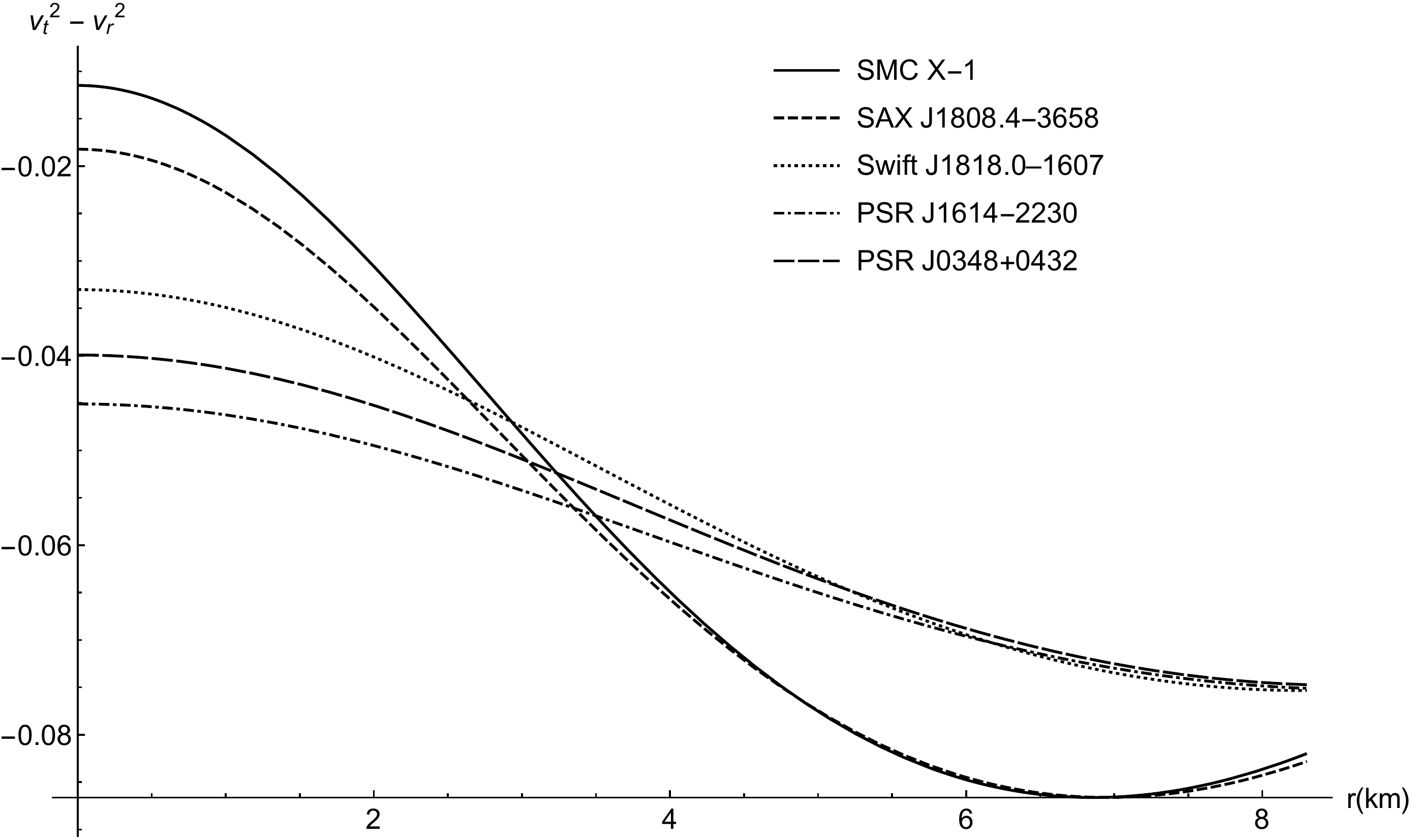}
	\caption{$\upsilon^2_t - \upsilon^2_r$ profile \label{fig13}}
\end{figure}

\begin{figure}[H]
	\centering
	\includegraphics[scale=0.5]{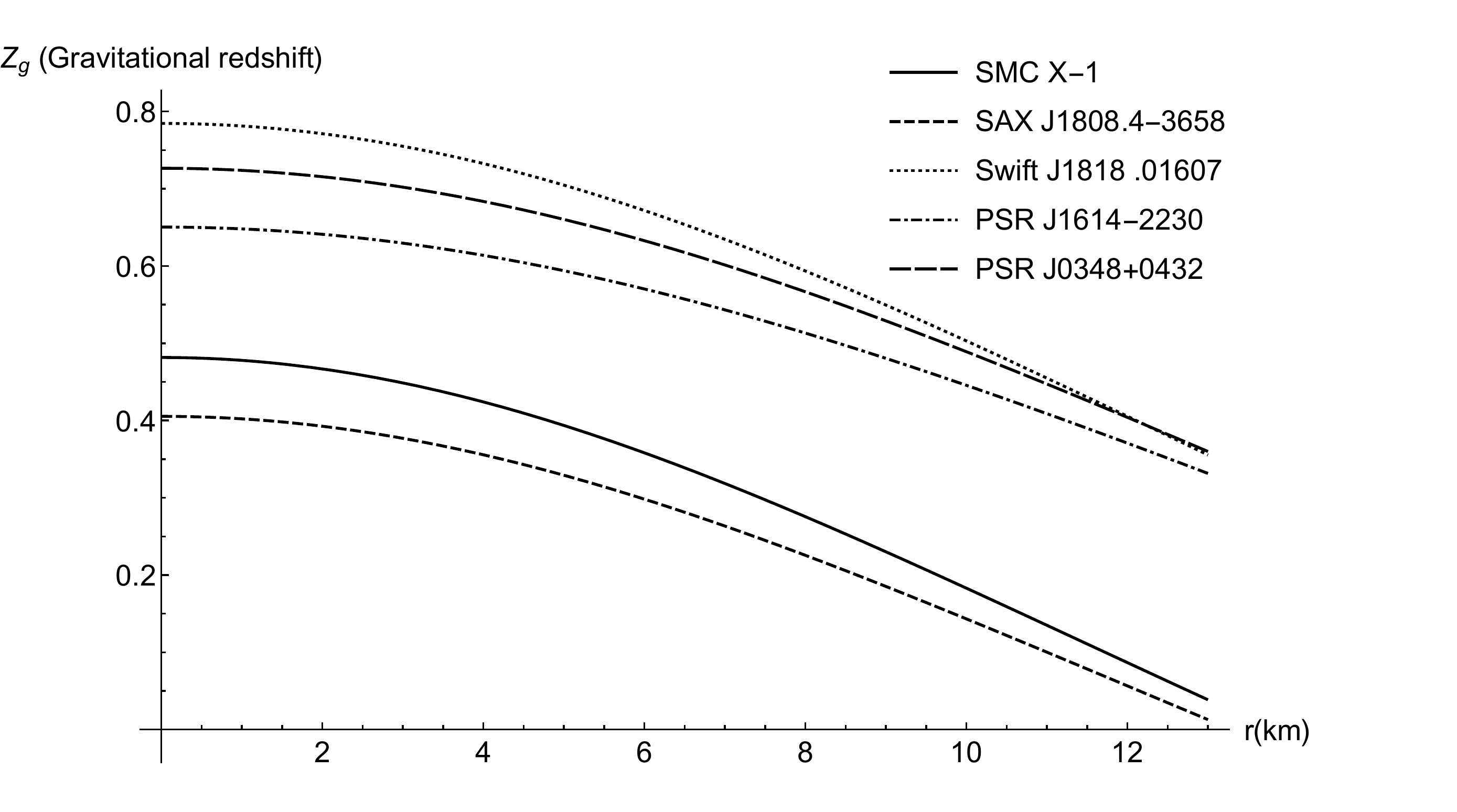}
	\caption{Gravitational Redshift \label{fig14}}
\end{figure}

\begin{figure}[H]
	\centering
	\includegraphics[scale=0.5]{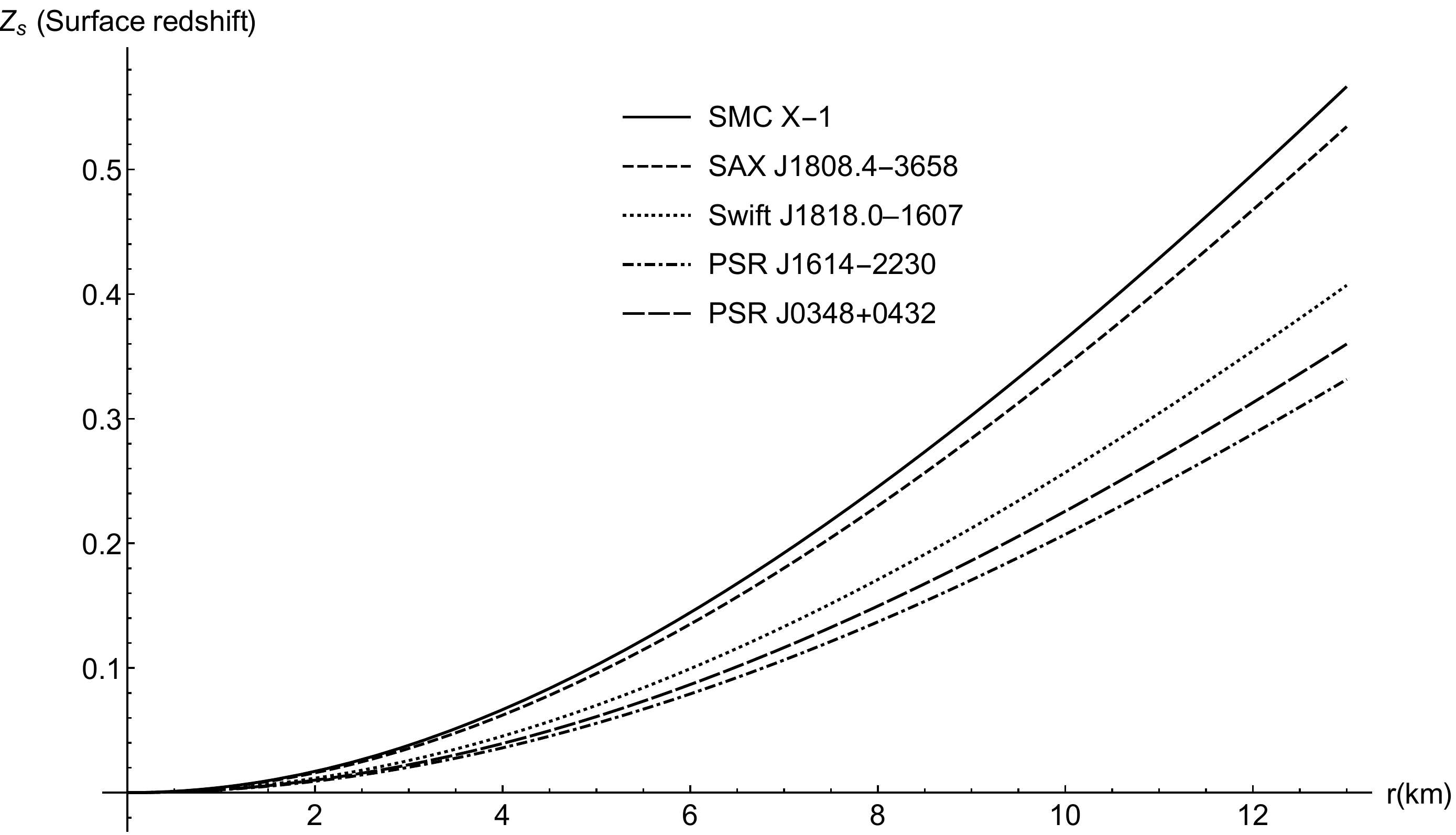}
	\caption{Surface Redshift \label{fig15}}
\end{figure}

\begin{figure}[H]
	\centering
	\includegraphics[scale=0.5]{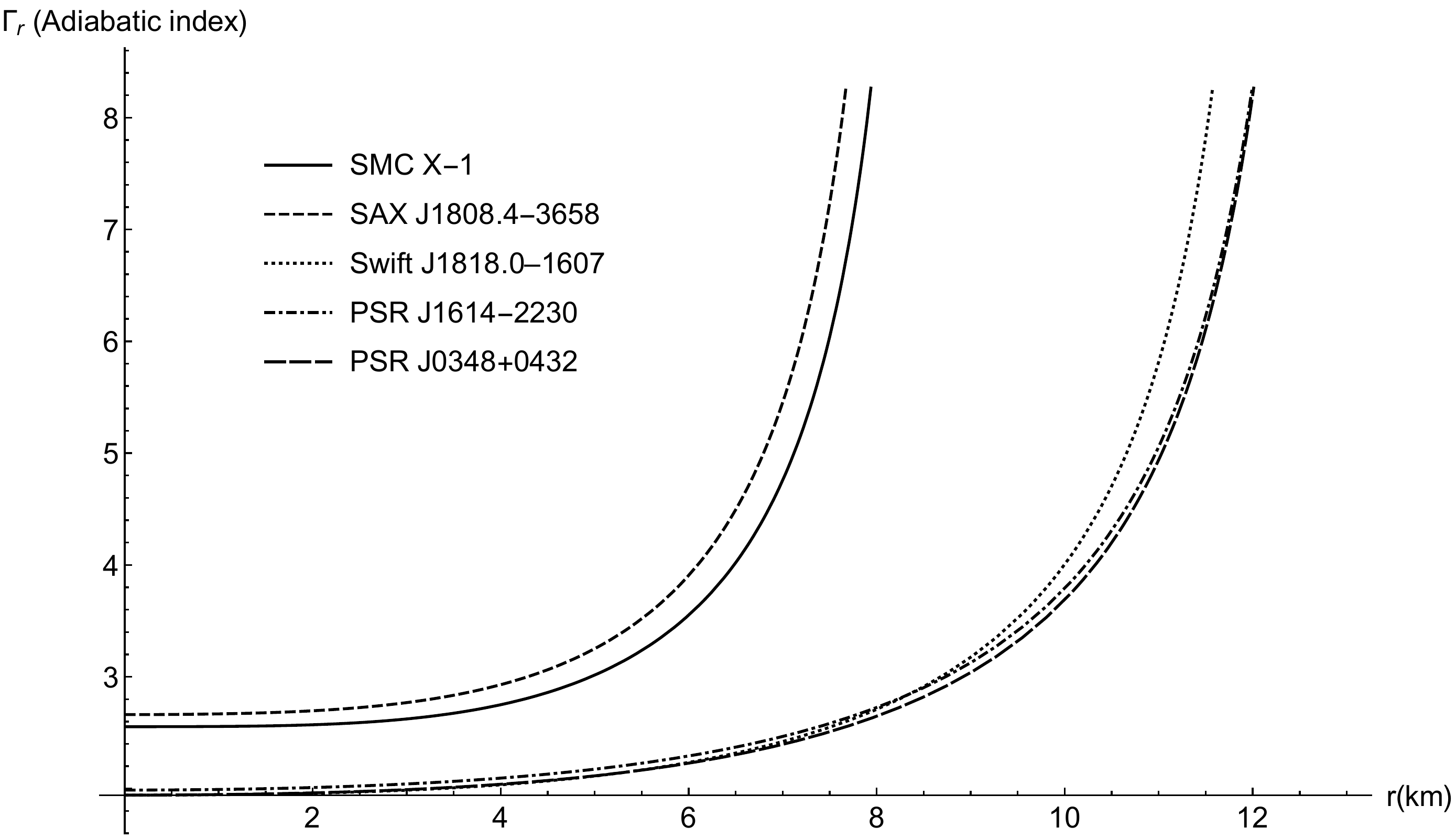}
	\caption{Adiabatic Index \label{fig16}}
\end{figure}

\begin{figure}[H]
	\centering
	\includegraphics[scale=0.35]{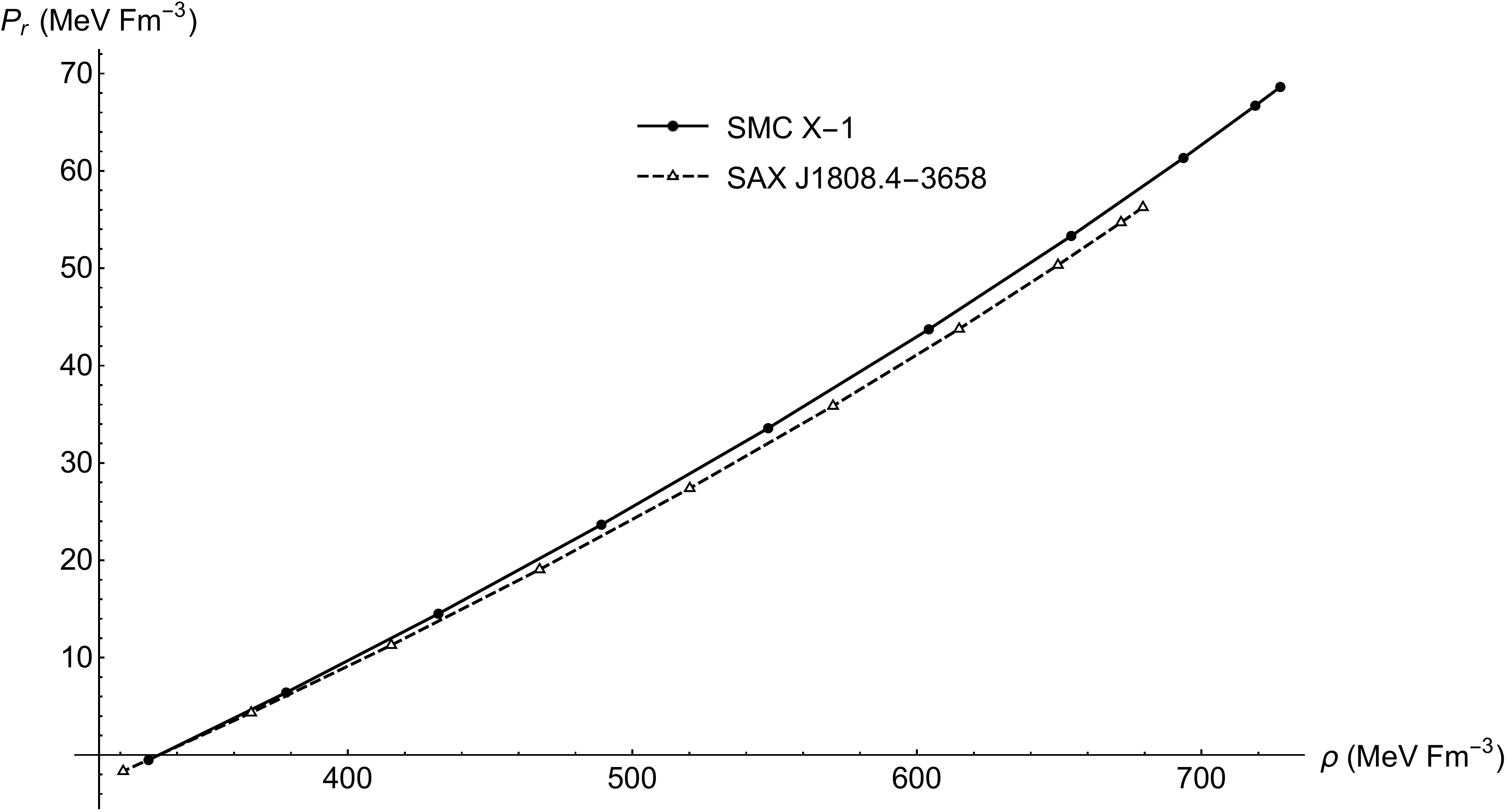}
	\caption{EOS for SMC X-1 and SAX J1808.4-3658\label{fig17}}
\end{figure}

\begin{figure}[H]
	\centering
	\includegraphics[scale=0.35]{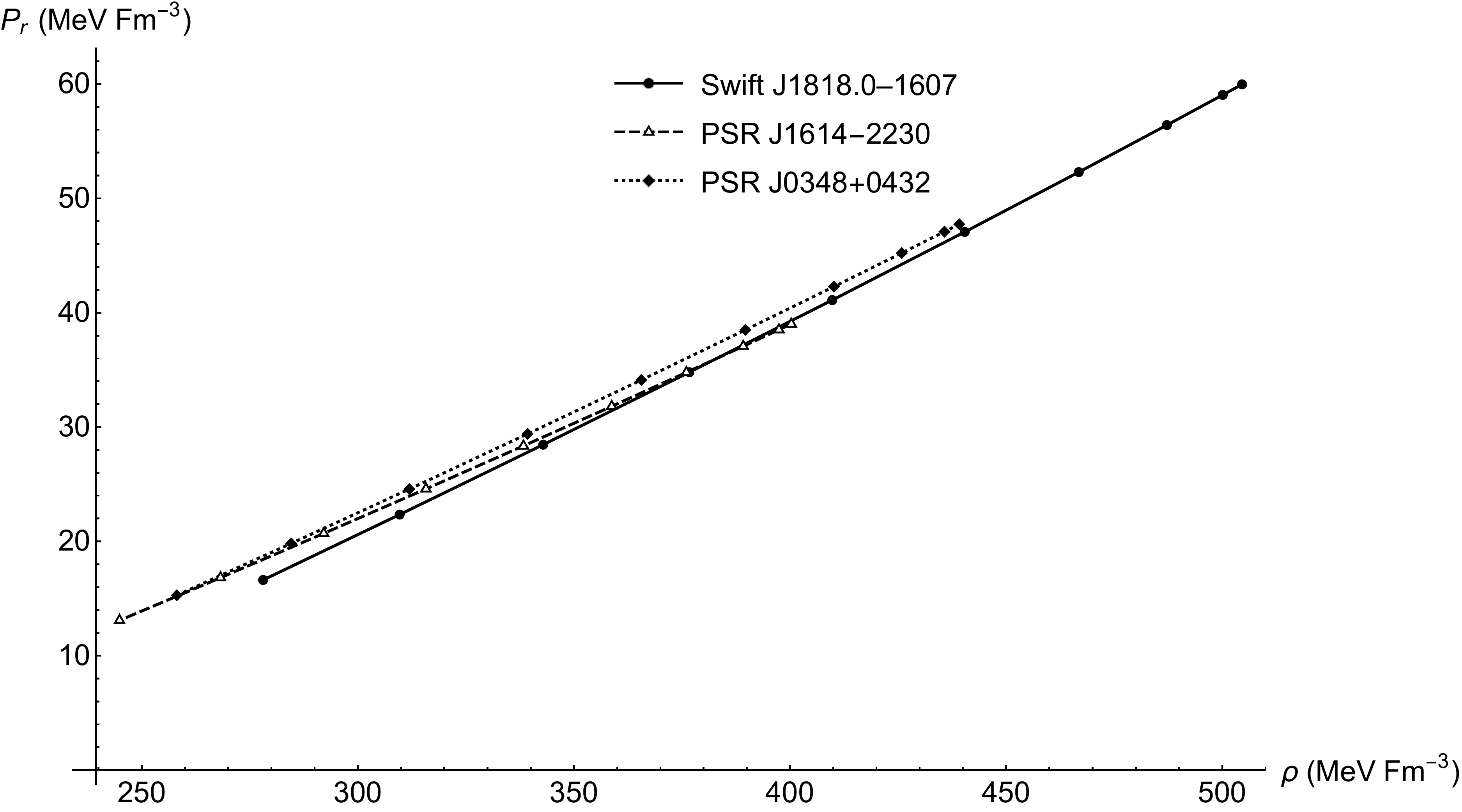}
	\caption{EOS for Swift J1818.0-1607, PSR J1614-2230, PSR J0348+0432 \label{fig18}}
\end{figure}

\begin{figure}[H]
	\centering
	\includegraphics[scale=0.5]{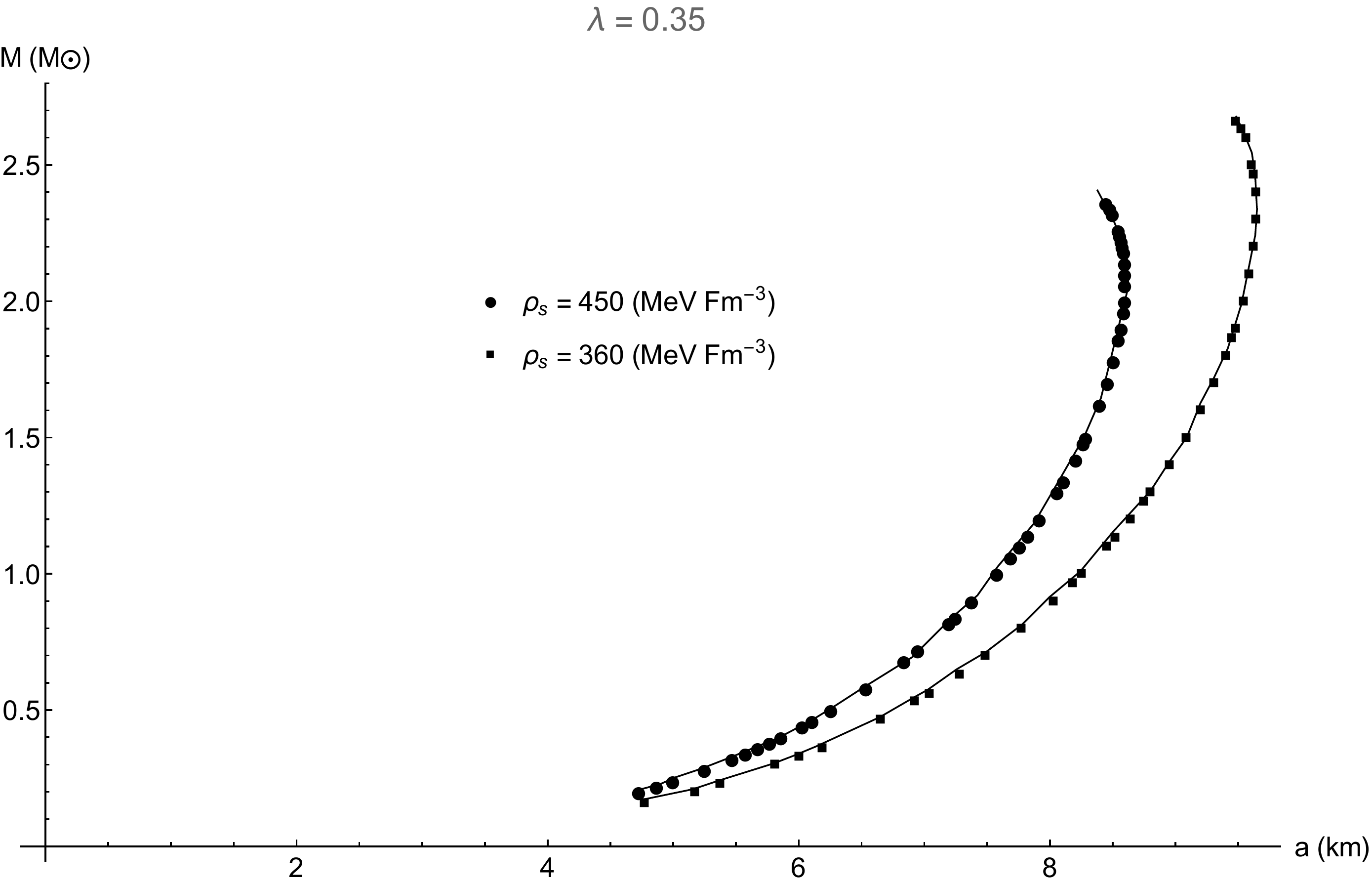}
	\caption{\textit{M-a} plot for $\lambda = 0.35$ at surface density of $\rho_s = 450$ and $\rho_s = 360 ~\text{MeV}\cdot \text{fm}^{-3}$\label{fig19}}
\end{figure}

\end{document}